\documentclass[twocolumn,pre,aps,nofootinbib]{revtex4}

\usepackage{graphicx}
\graphicspath{{./fig/}}

\usepackage{bm}

\begin{document}


%

\newcommand{\EQ}{\begin{equation}}
\newcommand{\EN}{\end{equation}}
\newcommand{\EQA}{\begin{eqnarray}}
\newcommand{\ENA}{\end{eqnarray}}
\newcommand{\eq}[1]{(\ref{#1})}
\newcommand{\EEq}[1]{Equation~(\ref{#1})}
\newcommand{\Eq}[1]{Eq.~(\ref{#1})}
\newcommand{\Eqs}[2]{Eqs~(\ref{#1}) and~(\ref{#2})}
\newcommand{\eqs}[2]{(\ref{#1}) and~(\ref{#2})}
\newcommand{\Eqss}[2]{Eqs~(\ref{#1})--(\ref{#2})}
\newcommand{\Sec}[1]{Sec.~\ref{#1}}
\newcommand{\Secs}[2]{Secs.~\ref{#1} and~\ref{#2}}
\newcommand{\Fig}[1]{Fig.~\ref{#1}}
\newcommand{\FFig}[1]{Figure~\ref{#1}}
\newcommand{\Tab}[1]{Table~\ref{#1}}
\newcommand{\Figs}[2]{Figures~\ref{#1} and \ref{#2}}
\newcommand{\Tabs}[2]{Tables~\ref{#1} and \ref{#2}}
\newcommand{\bra}[1]{\langle #1\rangle}
\newcommand{\bbra}[1]{\left\langle #1\right\rangle}
\newcommand{\mean}[1]{\overline #1}
\newcommand{\meanB}{\overline{B}}
\newcommand{\mod}[1]{\mid\!\!#1\!\!\mid}
\newcommand{\chk}[1]{[{\em check: #1}]}

\newcommand{\Reynolds}{\mathrm{Re}}
\newcommand{\Rm}{\mathrm{Re}_{\rm M}}
\newcommand{\Pm}{\mathrm{Pr}_{\rm M}}
\newcommand{\ii}{\mathrm{i}}

%
%
\newcommand{\gggg}{\bm{g}}
\newcommand{\ddd}{\bm{d}}
\newcommand{\rrr}{\bm{r}}
\newcommand{\xx}{\bm{x}}
\newcommand{\yy}{\bm{y}}
\newcommand{\zzz}{\bm{z}}
\newcommand{\uu}{\bm{u}}
\newcommand{\vv}{\bm{v}}
\newcommand{\ww}{\bm{w}}
\newcommand{\mm}{\bm{m}}
\newcommand{\PP}{\bm{P}}
\newcommand{\QQ}{\bm{Q}}
\newcommand{\UU}{\bm{U}}
\newcommand{\bb}{\bm{b}}
\newcommand{\qq}{\bm{q}}
\newcommand{\BB}{\bm{B}}
\newcommand{\HH}{\bm{H}}
\newcommand{\II}{\bm{I}}
\newcommand{\AAA}{\bm{A}}
\newcommand{\aaa}{\bm{a}}
\newcommand{\aaaa}{\bm{a}} 
\newcommand{\eee}{\bm{e}}
\newcommand{\jj}{\bm{j}}
\newcommand{\JJ}{\bm{J}}
\newcommand{\nn}{\bm{n}}
\newcommand{\ee}{\bm{e}}
\newcommand{\ff}{\bm{f}}
\newcommand{\EE}{\bm{E}}
\newcommand{\FF}{\bm{F}}
\newcommand{\TT}{\bm{T}}
\newcommand{\CC}{\bm{C}}
\newcommand{\KK}{\bm{K}}
\newcommand{\MM}{\bm{M}}
\newcommand{\GG}{\bm{G}}
\newcommand{\kk}{\bm{k}}
\newcommand{\SSS}{\bm{S}}
\newcommand{\grav}{\bm{g}}
\newcommand{\nab}{\bm{\nabla}}
\newcommand{\OO}{\bm{\Omega}}
\newcommand{\oo}{\bm{\omega}}
\newcommand{\LL}{\bm{\Lambda}}
\newcommand{\llambda}{\bm{\lambda}}
\newcommand{\pomega}{\bm{\varpi}}
%
%
\newcommand{\RRRR}{\bm{\mathsf{R}}}
\newcommand{\SSSS}{\bm{\mathsf{S}}}

\newcommand{\DD}{\mathrm{D}}
\newcommand{\dd}{\mathrm{d}}

\newcommand{\onethird}{{\textstyle{\frac{1}{3}}}}
\newcommand{\W}{\,{\rm W}}
\newcommand{\V}{\,{\rm V}}
\newcommand{\kV}{\,{\rm kV}}
\newcommand{\T}{\,{\rm T}}
\newcommand{\G}{\,{\rm G}}
\newcommand{\Hz}{\,{\rm Hz}}
\newcommand{\kHz}{\,{\rm kHz}}
\newcommand{\kG}{\,{\rm kG}}
\newcommand{\K}{\,{\rm K}}
\newcommand{\g}{\,{\rm g}}
\newcommand{\s}{\,{\rm s}}
\newcommand{\ms}{\,{\rm ms}}
\newcommand{\cm}{\,{\rm cm}}
\newcommand{\m}{\,{\rm m}}
\newcommand{\km}{\,{\rm km}}
\newcommand{\kms}{\,{\rm km/s}}
\newcommand{\kg}{\,{\rm kg}}
\newcommand{\Mm}{\,{\rm Mm}}
\newcommand{\pc}{\,{\rm pc}}
\newcommand{\kpc}{\,{\rm kpc}}
\newcommand{\yr}{\,{\rm yr}}
\newcommand{\Myr}{\,{\rm Myr}}
\newcommand{\Gyr}{\,{\rm Gyr}}
\newcommand{\erg}{\,{\rm erg}}
\newcommand{\mol}{\,{\rm mol}}
\newcommand{\dyn}{\,{\rm dyn}}
\newcommand{\J}{\,{\rm J}}
\newcommand{\RM}{\,{\rm RM}}
\newcommand{\EM}{\,{\rm EM}}
\newcommand{\AU}{\,{\rm AU}}
\newcommand{\A}{\,{\rm A}}
%
%
\newcommand{\yan}[3]{, Astron. Nachr. {\bf #2}, #3 (#1).}
\newcommand{\yact}[3]{, Acta Astron. {\bf #2}, #3 (#1).}
\newcommand{\yana}[3]{, Astron. Astrophys. {\bf #2}, #3 (#1).}
\newcommand{\yanas}[3]{, Astron. Astrophys. Suppl. {\bf #2}, #3 (#1).}
\newcommand{\yanal}[3]{, Astron. Astrophys. Lett. {\bf #2}, #3 (#1).}
\newcommand{\yass}[3]{, Astrophys. Spa. Sci. {\bf #2}, #3 (#1).}
\newcommand{\ysci}[3]{, Science {\bf #2}, #3 (#1).}
\newcommand{\ysph}[3]{, Solar Phys. {\bf #2}, #3 (#1).}
\newcommand{\yjetp}[3]{, Sov. Phys. JETP {\bf #2}, #3 (#1).}
\newcommand{\yspd}[3]{, Sov. Phys. Dokl. {\bf #2}, #3 (#1).}
\newcommand{\ysov}[3]{, Sov. Astron. {\bf #2}, #3 (#1).}
\newcommand{\ysovl}[3]{, Sov. Astron. Letters {\bf #2}, #3 (#1).}
\newcommand{\ymn}[3]{, Monthly Notices Roy. Astron. Soc. {\bf #2}, #3 (#1).}
\newcommand{\yqjras}[3]{, Quart. J. Roy. Astron. Soc. {\bf #2}, #3 (#1).}
\newcommand{\ynat}[3]{, Nature {\bf #2}, #3 (#1).}
\newcommand{\sjfm}[2]{, J. Fluid Mech., submitted (#1).}
\newcommand{\pjfm}[2]{, J. Fluid Mech., in press (#1).}
\newcommand{\yjfm}[3]{, J. Fluid Mech. {\bf #2}, #3 (#1).}
\newcommand{\ypepi}[3]{, Phys. Earth Planet. Int. {\bf #2}, #3 (#1).}
\newcommand{\ypr}[3]{, Phys.\ Rev.\ {\bf #2}, #3 (#1).}
\newcommand{\yprl}[3]{, Phys.\ Rev.\ Lett.\ {\bf #2}, #3 (#1).}
\newcommand{\yepl}[3]{, Europhys. Lett. {\bf #2}, #3 (#1).}
\newcommand{\pcsf}[2]{, Chaos, Solitons \& Fractals, in press (#1).}
\newcommand{\ycsf}[3]{, Chaos, Solitons \& Fractals{\bf #2}, #3 (#1).}
\newcommand{\yprs}[3]{, Proc. Roy. Soc. Lond. {\bf #2}, #3 (#1).}
\newcommand{\yptrs}[3]{, Phil. Trans. Roy. Soc. {\bf #2}, #3 (#1).}
\newcommand{\yjcp}[3]{, J. Comp. Phys. {\bf #2}, #3 (#1).}
\newcommand{\yjgr}[3]{, J. Geophys. Res. {\bf #2}, #3 (#1).}
\newcommand{\ygrl}[3]{, Geophys. Res. Lett. {\bf #2}, #3 (#1).}
\newcommand{\yobs}[3]{, Observatory {\bf #2}, #3 (#1).}
\newcommand{\yaj}[3]{, Astronom. J. {\bf #2}, #3 (#1).}
\newcommand{\yapj}[3]{, Astrophys. J. {\bf #2}, #3 (#1).}
\newcommand{\yapjs}[3]{, Astrophys. J. Suppl. {\bf #2}, #3 (#1).}
\newcommand{\yapjl}[3]{, Astrophys. J. {\bf #2}, #3 (#1).}
\newcommand{\ypp}[3]{, Phys. Plasmas {\bf #2}, #3 (#1).}
\newcommand{\ypasj}[3]{, Publ. Astron. Soc. Japan {\bf #2}, #3 (#1).}
\newcommand{\ypac}[3]{, Publ. Astron. Soc. Pacific {\bf #2}, #3 (#1).}
\newcommand{\yannr}[3]{, Ann. Rev. Astron. Astrophys. {\bf #2}, #3 (#1).}
\newcommand{\yanar}[3]{, Astron. Astrophys. Rev. {\bf #2}, #3 (#1).}
\newcommand{\yanf}[3]{, Ann. Rev. Fluid Dyn. {\bf #2}, #3 (#1).}
\newcommand{\ypf}[3]{, Phys. Fluids {\bf #2}, #3 (#1).}
\newcommand{\yphy}[3]{, Physica {\bf #2}, #3 (#1).}
\newcommand{\ygafd}[3]{, Geophys. Astrophys. Fluid Dynam. {\bf #2}, #3 (#1).}
\newcommand{\yzfa}[3]{, Zeitschr. f. Astrophys. {\bf #2}, #3 (#1).}
\newcommand{\yptp}[3]{, Progr. Theor. Phys. {\bf #2}, #3 (#1).}
\newcommand{\yjour}[4]{, #2 {\bf #3}, #4 (#1).}
\newcommand{\pjour}[3]{, #2, in press (#1).}
\newcommand{\sjour}[3]{, #2, submitted (#1).}
\newcommand{\yprep}[2]{, #2, preprint (#1).}
\newcommand{\pproc}[3]{, (ed. #2), #3 (#1) (to appear).}
\newcommand{\yproc}[4]{, (ed. #3), pp. #2. #4 (#1).}
\newcommand{\ybook}[3]{, {\em #2}. #3 (#1).}

\preprint{NORDITA 2003-42 AP}

\title{Simulations of nonhelical hydromagnetic turbulence}

\author{Nils Erland L.\ Haugen}
  \affiliation{Department of Physics, The Norwegian University of Science
  and Technology, H{\o}yskoleringen 5, N-7034 Trondheim, Norway}
  \email{nils.haugen@phys.ntnu.no}
\author{Axel Brandenburg}
  \affiliation{NORDITA, Blegdamsvej 17, DK-2100 Copenhagen \O, Denmark}
  \email{brandenb@nordita.dk}
\author{Wolfgang Dobler}
  \email{Wolfgang.Dobler@kis.uni-freiburg.de}
  \affiliation{Kiepenheuer-Institut f\"ur Sonnenphysik,
  Sch\"oneckstra\ss{}e 6, D-79104 Freiburg, Germany}

\date{\today,~ $ $Revision: 1.184 $ $}

\begin{abstract}
Nonhelical hydromagnetic forced turbulence is investigated using large scale
simulations on up to $256$ processors and $1024^3$ meshpoints.
The magnetic Prandtl number is varied between 1/8 and 30, although
in most cases it is unity.
When the magnetic Reynolds number is based on the inverse forcing
wavenumber, the critical value for dynamo action is shown to be around
35 for magnetic Prandtl number of unity.
For small magnetic Prandtl numbers we find the critical magnetic Reynolds
number to increase with decreasing magnetic Prandtl number.
The Kazantsev $k^{3/2}$ spectrum for magnetic energy is confirmed
for the kinematic regime, i.e.\ when nonlinear effects are still unimportant
and when the magnetic Prandtl number is unity.
In the nonlinear regime,
the energy budget converges for large Reynolds numbers (around 1000) such
that for our parameters
about 70\% is in kinetic energy and about 30\% is in magnetic energy.
The energy dissipation rates are converged to 30\% viscous dissipation
and 70\% resistive dissipation.
Second order structure functions of the Elsasser variables give evidence for
a $k^{-5/3}$ spectrum.
Nevertheless, the three-dimensional spectrum is close to $k^{-3/2}$,
but we argue that this is due to the bottleneck effect.
The bottleneck effect is shown to be equally strong both for magnetic and nonmagnetic
turbulence, but it is far weaker in one-dimensional spectra that are
normally studied in laboratory turbulence.
Structure function exponents for other orders are well described by
the She-Leveque formula, but the velocity field is significantly
less intermittent and the magnetic field is more intermittent than the 
Elsasser variables.
\end{abstract}
\maketitle

\section{Introduction}

Dynamo action, i.e.\ the conversion of kinetic energy into magnetic
energy, plays an important role in astrophysical bodies ranging
from stars to galaxies and even clusters of galaxies. The gas in these
bodies is turbulent and the magnetic Reynolds numbers are huge ($10^{10}$
to $10^{20}$). This suggests that there should be dynamo action and that
the magnetic fields should be amplified on the time scale of the
turbulent turnover time \cite{Bat50,BS51}.
The magnetic fields of many astrophysical bodies show a great deal of
spatio-temporal order (e.g.~an 11-year cycle and equatorward migration of the
magnetic field in the case of the sun),
which can basically be explained by the helicity effect
\cite{SKR66,PFL76,B01}.
In many other astrophysical environments, however, the helicity effect is
probably completely irrelevant (e.g.\ in the solar wind
\cite{GRF94}, clusters of galaxies \cite{RSB99}, and the early universe
after recombination \cite{KCOR97}).
In all these cases the magnetic field does not show spatio-temporal order
of the type known for helical hydromagnetic turbulence \cite{B01}.
Both analytic theory \cite{Kaz68} as well as simulations
\cite{MFP81} have long shown that dynamo action is possible even
without kinetic helicity, but that the field is then spatially highly
intermittent with substantial power at small length scales.
Hydromagnetic turbulence has recently also been studied in the laboratory
\cite{LabTurb,Forest,Lathrop},
but there the magnetic Reynolds numbers are still rather small.
Therefore, numerical simulations are currently the most powerful tool.

In spite of significant progress over the past two decades,
the form of the energy spectrum at large magnetic
Reynolds numbers is still a matter of debate.
Particular progress has been made in the
case where there is a large scale field. Goldreich \& Sridhar \cite{GS95}
have proposed that the magnetic energy spectrum has a $k^{-5/3}$ 
inertial range due to the anisotropy imposed by the local magnetic field.
This was in conflict with the earlier Iroshnikov-Kraichnan \cite{Kra65}
$k^{-3/2}$ spectrum, which
assumes isotropy, but can now be ruled out \cite{CV00,BM00,MG01}.
Some simulations still show a $k^{-3/2}$ spectrum, but this is probably
due to the bottleneck effect; see below.
A $k^{-5/3}$ total energy spectrum has however clearly been
seen in decaying hydromagnetic turbulence \cite{BM00}.
Here the total energy spectrum is simply the sum of kinetic and magnetic
energy spectra.

The case of an imposed large scale magnetic field
with approximate equipartition strength is in some respects
similar to the case of a self-generated large scale field which emerges
if there is helicity in the flow \cite{MFP81,PFL76,BP99}. In the latter
case of helical turbulence,
and for unit magnetic Prandtl number, kinetic and magnetic energy spectra
are in almost perfect equipartition on all scales smaller than the forcing
scale (see Fig.~11 of Ref.~\cite{B01}), and
the two spectra tend to approach a $k^{-5/3}$ inertial range as the
magnetic Reynolds number is increased.
In the former case, equipartition throughout the inertial range may
require that the energy of the imposed field is comparable to the
kinetic energy; for stronger fields the magnetic energy becomes suppressed
at small scales \cite{HB04}.
If the magnetic Prandtl number is larger than unity, the resistive
cutoff is prolonged to larger wavenumbers by a $k^{-1}$ spectrum
\cite{CLV02b}.
Nevertheless some authors have suggested that even in the nonlinear regime
the spectral magnetic energy increases toward smaller scales, similar to the
kinematic regime where the energy spectrum scales as $k^{+3/2}$
\cite{Kaz68,KA92}.

In the absence of an imposed large scale magnetic field, and with no
helicity, the situation is in many ways different. Early 
simulations suggest that the magnetic field is dominated by
small scale power \cite{MFP81}.
Even for a magnetic Prandtl number of unity the magnetic energy exceeds
the kinetic energy at small scales \cite{KYM91,Chou01,MC01}, a result
that is otherwise (with imposed field or with helicity)
only obtained for magnetic Prandtl numbers larger than
unity \cite{MC01,CLV02b,B01,SCHMMW02}.

The main problem in determining the energy spectrum at
large hydrodynamic and magnetic Reynolds numbers
is the lack of a proper inertial range in the magnetic field. Another
problem is that in the absence of helicity the dynamo is much weaker and
one needs significantly (about $20\times$) larger magnetic Reynolds
numbers before the dynamo is even excited. In practice, this means that
a resolution of $128^3$ or more is mandatory.

The shortness or even lack of an inertial range has frequently led to the
use of hyperviscosity and hyperdiffusivity \cite{MFP81,KYM91},
which has the tendency to
extend the inertial range and to shorten the dissipative subrange.
However, in recent years it has become clear that these
modifications to the viscosity and diffusion operators can affect major
parts of the inertial subrange and make it shallower. This is also
referred to as the bottleneck effect which is present already for
ordinary viscosity \cite{Fal94}, but it becomes greatly exaggerated with
hyperviscosity; compare Figs 8 and 11 of Ref.~\cite{BM00}.
In helical dynamos the saturation time becomes significantly
prolonged and the final saturation level is artificially enhanced
\cite{BS02}. It is therefore important to use simulations with regular
viscosity and magnetic diffusivity at high enough resolution.

A lot of work has already been done in order to determine the form of
the magnetic energy spectrum at large magnetic Reynolds numbers.
Some of the relevant papers are listed in \Tab{Tsum}, where we indicate
the main properties of their set-ups.
In addition to the papers mentioned above we have included a number
of additional ones \cite{MMDM01,OMG98,CL03,MLKB98,CLV02,BNP02,HBD03},
some of which will be discussed below in more detail.

\begingroup
\begin{table}[t!]
\centering
\caption{
Summary of earlier work on hydromagnetic turbulence simulations.
In addition to the reference of each paper we give
for convenience also a more descriptive abbreviation.
In Ref.~\cite{MG01} the largest resolution was $N^3=256^2\times512$,
so we have listed the geometrical mean corresponding to $322^3$.
The column `hyper' indicates whether or not hyperviscosity
or hyperdiffusivity have been used; `artif' stands for shock-capturing
artificial viscosity.
In the column `forced', the abbreviation `hel' stands for helical, `noh'
for non-helical,
'decay' for decaying turbulence without forcing, and
`$\sim$noh' means that there is some small fraction of helicity, but due
to a small scale separation between the forcing and the box scale the 
simulation is practically non-helical.
In the column $\left<\BB\right>$, `ext' indicates the use of an external
field.
The column `compr' signifies whether the models are compressible.
}
\label{Tsum}
\begin{ruledtabular}
\begin{tabular}{lrrlcclc}
Paper &  Ref.\      & $N$ & hyper & forced &$\bra{\BB}$& compr & $\Pm$ \\
\hline
MFP81 & \cite{MFP81}&  64 &  both &hel/noh  &   0   &   no   & 1 \\
B01   & \cite{B01}  & 120 &  no   & hel     &   0   &  yes   & 0.1-100\\
KYM91 & \cite{KYM91}& 128 &  yes  &  hel    &   0   &   no   & 1 \\
MMDM01& \cite{MMDM01}&128 &  no   & decay   & ext/0 &   no   & 1 \\
OMG98 &\cite{OMG98} & 128 &  no   &  decay  & ext/0 &  both  & 1 \\
CL03  & \cite{CL03} & 216 &       &  decay  & ext   &  yes   & 1 \\
MLKB98&\cite{MLKB98}& 256 & artif &  decay  &    0  &  yes   & 1 \\
CV00  & \cite{CV00} & 256 &  both &$\sim$noh&  ext  &   no   & 1 \\
MC01  & \cite{MC01} & 256 &  both &  noh    & ext/0 &   no   & 1-2500\\
MG01  & \cite{MG01} & 322 &  yes  &  noh    &  ext  &   no   & 1 \\
CLV02 & \cite{CLV02}& 384 &  yes  &$\sim$noh&  ext  &   no   & large \\
BNP02 & \cite{BNP02}& 500 &  yes  &  noh    &  ext  &  yes   & 1 \\
BM00  & \cite{BM00} & 512 &  both &  decay  &    0  &   no   & 1 \\
HBD03 & \cite{HBD03}&1024 &   no  &  noh    &    0  &  yes   & 1 \\
\end{tabular}
\end{ruledtabular}
\end{table}
\endgroup

In a recent paper we have already presented some initial
results on energy spectra at a resolution of up to
$1024^3$ meshpoints; see Ref.~\cite{HBD03}, hereafter referred to as Paper~I.
In the present paper we discuss the associated results for the
critical magnetic Reynolds number for dynamo action, we consider
a range of different values of the magnetic Prandtl number,
compare with the case of finite magnetic helicity, and
present visualizations of the magnetic field and the dissipative structures in
hydromagnetic turbulence. We look in detail at the energy spectra and the
structure functions of total, kinetic and magnetic energies. 
The simulations have been carried out using the Pencil Code
\cite{PencilCode} which is a
memory-efficient high-order finite difference code using the $2N$-RK3
scheme of Williamson \cite{Wil80}.
Our approach is technically similar to that used in Ref.~\cite{B01}.
The Pencil Code is fully compressible; we therefore consider the weakly
compressible case (the Mach number is around 0.1), which can be considered
as an approximation to the incompressible case.
A list of parameters for most runs discussed in the present paper is given
in Table~\ref{Truns}.

\begin{table}[H!tbp]
\caption{
Summary of runs with $\Pm=1$ (thus $\Reynolds=\Rm$) and forcing at
$k_{\rm f}=1.5$.
In all cases, except D2, we have $u_{\rm rms}\approx0.12$. Run D2 had a 
stronger forcing, and therefore $u_{\rm rms}\approx 0.18$. The
Taylor microscale Reynolds number is $\Reynolds_{\lambda}=
u_{\rm 1D}\ell_\lambda/\nu$, where $u_{\rm 1D}=u_{\rm rms}/\sqrt{3}$
is the rms velocity in one direction and
$\ell_\lambda=\sqrt{5}u_{\rm rms}/\omega_{\rm rms}$ is the Taylor microscale,
where $\omega_{\rm rms}$ is the rms vorticity.
\label{Truns}}
\begin{ruledtabular}
\begin{tabular}{cccccccc}
Run&Resolution&$\nu\times 10^{4}$&$\Reynolds_{\lambda}$&$\Rm$&$B_{\rm rms}$
&$\epsilon_{\rm K}\times 10^{4}$&$\epsilon_{\rm M}\times 10^{4}$     \\
\hline
A  & $64^3$  &$7.0$&  80 &    120    &      0.052  &$1.0$ &$1.2$\\ 
B  &$128^3$  &$4.0$& 110 &    190    &      0.060  &$0.79$&$1.4$\\ 
C  &$256^3$  &$2.0$& 160 &    420    &      0.062  &$0.78$&$1.6$\\ 
D  &$512^3$  &$1.5$& 190 &    540    &      0.072  &$0.68$&$1.7$\\ 
D2 &$512^3$  &$2.0$& 180 &    600    &      0.092  &$2.2$ &$5.0$\\ 
E  &$1024^3$ &$0.8$& 230 &    960    &      0.075  &$0.63$&$1.5$\\ 
\end{tabular}
\end{ruledtabular}
\end{table}

\section{Equations}

We adopt an
isothermal equation of state with constant (isothermal) sound speed $c_{\rm s}$,
so the pressure $p$ is related to the density $\rho$ by
$p=\rho c_{\rm s}^2$. The equation of motion is written in the form
\EQ
{\DD\uu\over\DD t}=-c_{\rm s}^2\nab\ln\rho+{\JJ\times\BB\over\rho}
+\FF_{\rm visc}+\ff,
\label{dudt}
\EN
where $\DD/\DD t=\partial/\partial t+\uu\cdot\nab$ is the advective
derivative, $\JJ=\nab\times\BB/\mu_0$ is the current density, $\mu_0$
is the vacuum permeability,
\EQ
\FF_{\rm visc}=\nu\left(\nabla^2\uu+\onethird\nab\nab\cdot\uu
+2\SSSS\cdot\nab\ln\rho\right)
\EN
is the viscous force,
$\nu=\text{const}$ is the kinematic viscosity,
\EQ
{\sf S}_{ij}=\frac{1}{2}\left({\partial u_i\over\partial x_j}
+ {\partial u_j\over\partial x_i}
-\frac{2}{3} \delta_{ij}\nab\cdot\uu\right)
\EN
is the traceless rate of strain tensor, and $\ff$ is a
random forcing function that consists of non-helical
plane waves (see below). The continuity equation is
written in terms of the logarithmic density,
\EQ
{\DD\ln\rho\over\DD t}=-\nab\cdot\uu,
\EN
and the induction equation is solved in terms of the magnetic vector
potential $\AAA$, with $\BB=\nab\times\AAA$, so
\EQ
{\partial\AAA\over\partial t}=\uu\times\BB+\eta\nabla^2\AAA,
\label{dAdt}
\EN
where $\eta=\text{const}$ is the magnetic diffusivity.

We use periodic boundary conditions in all three directions
for all variables. This implies that the mass in the box
is conserved, i.e.\ $\bra\rho=\rho_0$, where $\rho_0$ is the value
of the initially uniform density, and angular brackets denote
volume averages. We adopt a forcing function $\ff$ of the form
\EQ
\ff(\xx,t)=\Reynolds\{N\ff_{\kk(t)}\exp[\ii\kk(t)\cdot\xx+\ii\phi(t)]\},
\EN
where
$\xx$ is the position vector.
The wave vector $\kk(t)$ and the random phase
$-\pi<\phi(t)\le\pi$ change at every time step, so $\ff(\xx,t)$ is
$\delta$-correlated in time.
For the time-integrated forcing function to be independent
of the length of the time step $\delta t$, the normalization factor $N$
has to be proportional to $\delta t^{-1/2}$.
On dimensional grounds it is chosen to be
$N=f_0 c_{\rm s}(|\kk|c_{\rm s}/\delta t)^{1/2}$, where $f_0$ is a
nondimensional forcing amplitude.
The value of the coefficient $f_0$ is chosen such that the maximum Mach
number stays below about 0.5; in practice this means $f_0=0.02\ldots0.05$.

At each timestep we select randomly one of many possible wavevectors
in a certain range
around a given forcing wavenumber.
The average wavenumber is referred to as $k_{\rm f}$.
We force the system with nonhelical transversal waves,
\EQ
\ff_{\kk}=\left(\kk\times\eee\right)/\sqrt{\kk^2-(\kk\cdot\eee)^2},
\label{nohel_forcing}
\EN
where $\eee$ is an arbitrary unit vector not aligned with $\kk$;
note that $|\ff_{\kk}|^2=1$.

The resulting flows are characterized by the kinetic and magnetic Reynolds
numbers,
\EQ
\Reynolds={u_{\rm rms}\over\nu k_{\rm f}},\quad
\Rm={u_{\rm rms}\over\eta k_{\rm f}},
\EN
respectively.
Their ratio is the magnetic Prandtl number,
\EQ
\Pm=\nu/\eta=\Rm/\Reynolds,
\EN
which is unity for most of the runs.

We use non-dimensional quantities by measuring length in units of
$1/k_1$ (where $k_1=2\pi/L$ is the smallest wavenumber in the box of size
$L$),
speed in units of the isothermal sound speed $c_{\rm s}$, density
in units of the initial value $\rho_0$, and magnetic field in
units of $(\mu_0\rho_0 c_{\rm s}^2)^{1/2}$.

\section{The kinematic phase and approach to saturation}

As initial condition we use a weak random seed magnetic field.
In this section we consider the time interval during which the
amplitude of the magnetic energy spectrum grows exponentially
and is still small compared to the kinetic energy spectrum
for all wavenumbers.

\subsection{Critical magnetic Reynolds number}

\begin{figure}[t!]\centering\includegraphics[width=0.45\textwidth]
{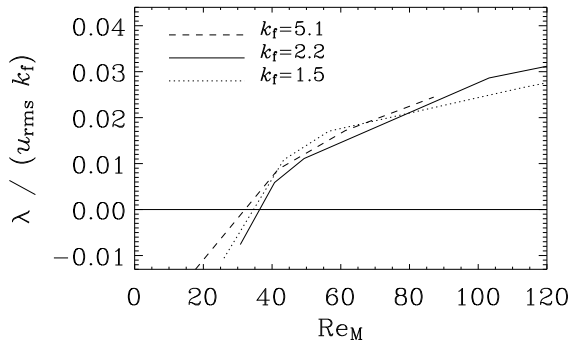}\caption{
Growth rate versus $\Rm$ for different values of $k_{\rm f}$, and
for $\Pm=1$.
The curves represent linear fits through the data.
Note that the critical value is around 35 for all the different runs.
The resolution varies between $64^3$ and $256^3$, and
$f_0=0.05$ in all runs, resulting in $u_{\rm rms}\approx0.2$.
}\label{Fpgrowth}\end{figure}

We have determined the growth rate of the rms magnetic field,
$\lambda\equiv\dd\ln B_{\rm rms}/ \dd t$,
for different values of $\Rm$ and $\Pm$.
Interpolating the curves $\lambda=\lambda(\Rm)$ through zero,
we find the critical value, $\Rm^{\rm(crit)}$,
which is roughly independent of $k_{\rm f}$; see \Fig{Fpgrowth}.
For the case $\Pm=1$ we find $\Rm^{\rm(crit)} \approx 35$, 
which is consistent
with the value obtained using a modified version of the
eddy-damped quasi-normal Markovian (EDQNM)
approximation \cite{LPF81} which gives $\Rm^{\rm(crit)}=29$.

\begin{figure}[t!]\centering\includegraphics[width=0.45\textwidth]
{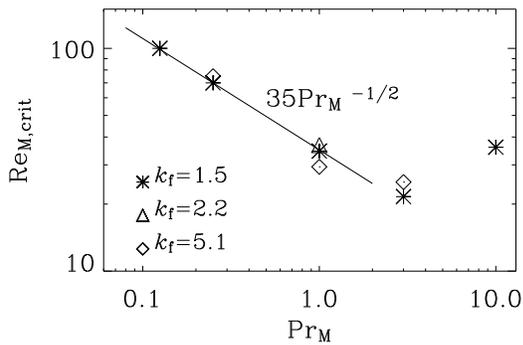}\caption{
Critical magnetic Reynolds number as a function of magnetic Prandtl number
for runs with different forcing scale.
}\label{prmcrit}\end{figure}

\begin{figure}[t!]\centering\includegraphics[width=0.45\textwidth]
{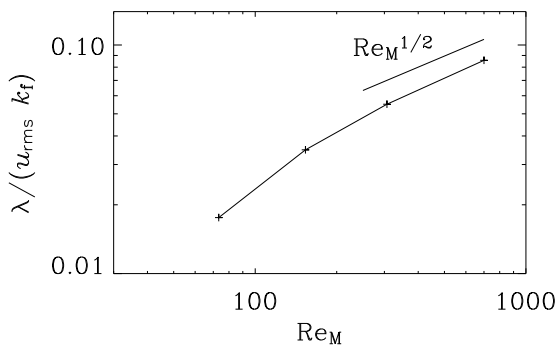}\caption{
Log-log plot of growth rate versus $\Rm$ for $k_{\rm f}=1.5$, and $\Pm=1$.
Note that $\lambda/(u_{\rm rms}k_{\rm f})$ seems to approach
$\Rm^{1/2}/300$ for large values of $\Rm$.
The resolution varies between $128^3$ and $512^3$.
}\label{Fpgrowth_new}\end{figure}
 
As the value of $\Pm$ is lowered, the critical magnetic Reynolds number
increases slowly like
\EQ
\Rm^{\rm(crit)}\approx35\,\Pm^{-1/2},
\EN
see \Fig{prmcrit}.
It remains uncertain whether this scaling persists to very small
values of $\Pm$ that are relevant to liquid metal experiments
or to stellar convection zones.
We note, however, that the EDQNM approximation predicts \cite{LPF81}
that the critical magnetic Reynolds number
is independent of $\Pm$ for a large range of $\Reynolds$ and $\Pm$.
Based on application of the Kazantsev model,
Schekochihin et al.\ \cite{Scheko04} have shown that,
if the correlation time is assumed to be independent of wavenumber,
there exists a finite value of $\Pm$ below which dynamo action
is impossible.
On the other hand, if the correlation time is proportional to
the eddy turnover time, $\sim k^{-2/3}$, dynamo action should
be possible even for very small values of $\Pm$.

For larger magnetic Reynolds numbers the growth rate of the magnetic
field approaches a $\Rm^{1/2}$ dependence; see \Fig{Fpgrowth_new}.
Such a power law would be expected if the growth rate is proportional
to the eddy turnover time at the dissipation wavenumber, $k_{\rm d}$,
i.e.\ $\lambda\propto k_{\rm d}^{2/3}\propto\Rm^{1/2}$, where we have
used $\Rm\propto k_{\rm d}^{4/3}$ \cite{Spriv}.
We must emphasize, however, that 
the $\Rm$ scaling is as yet rather short.

\subsection{The Kazantsev spectrum}
\label{SKazantsev}

Under the somewhat unrealistic assumption that the velocity field is
$\delta$-correlated in time, one can, for a given spatial correlation
function of the velocity, and ignoring magnetic feedback, derive an
evolution equation for the correlation function of the magnetic field
\cite{Kaz68,Nov83} (see Ref.~\cite{Sub99} for a nonlinear extension).
This can be rewritten as an integro-differential equation in $k$-space
which, in turn, can be written as a diffusion equation in $k$-space
if the velocity field has only power at large scales \cite{Kaz68,KA92}.
The result is $E^{\rm M}_k(k)\propto k^{3/2}K_{n(\lambda)}(k/k_\eta)$,
where $K_n$ is the Macdonald function
(modified Bessel function of the third kind),
$\lambda\approx3/4$ is an eigenvalue, and $n(\lambda)\approx0$;
see Ref.~\cite{Schek02} for details.
In linear theory the amplitude of the solution is however undetermined
and grows exponentially if the magnetic Reynolds number exceeds a
critical value of about 30-60 \cite{Nov83}.

\begin{figure}[t!]\centering\includegraphics[width=0.5\textwidth]
{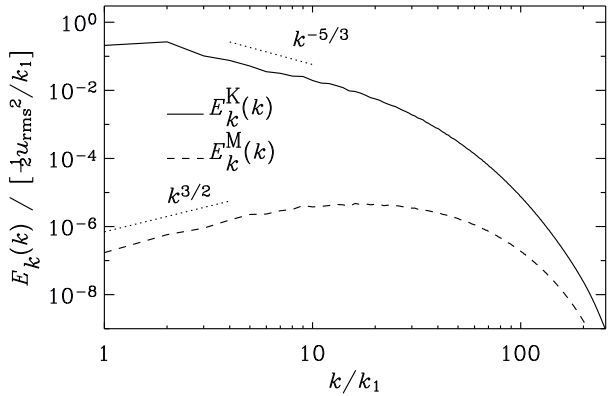}\caption{
Early spectra of kinetic and magnetic energy,
normalized by $\frac{1}{2}u_{\rm rms}^2/k_1$, 
during the kinematic stage of run D2.
}\label{plot_poweru_early}\end{figure}

\begin{figure}[t!]\centering\includegraphics[width=0.5\textwidth]
{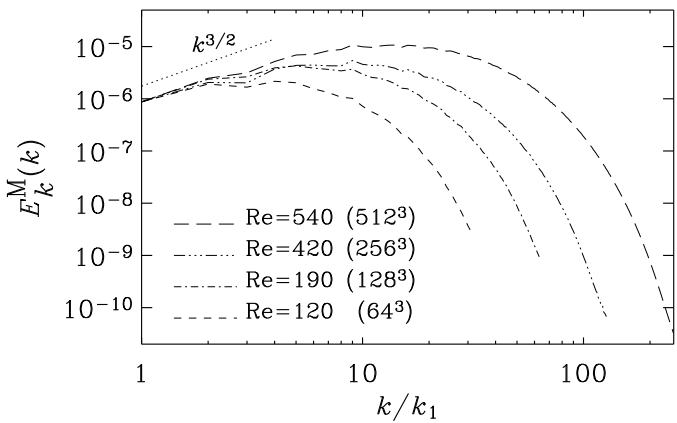}\caption{
Convergence of the
magnetic power spectrum toward the $k^{3/2}$ scaling
as $\Rm$ increases.
All spectra correspond to the
kinematic stages of Runs~A-D with $\Pm=1$,
but have been rescaled to make them coincide at small values of $k$.
The dimension on the ordinate is therefore arbitrary.
}\label{mag32_comp}\end{figure}

In comparison with our simulations we see qualitative agreement as
far as the $k^{3/2}$ slope is concerned.
As long as the magnetic field is weak, the velocity field has an
energy spectrum with the expected $k^{-5/3}$ Kolmogorov scaling; see
Fig \ref{plot_poweru_early}. During this phase the spectral magnetic energy
grows at all wavenumbers exponentially in time and the spectrum 
has the expected $k^{3/2}$ Kazantsev \cite{Kaz68} slope; 
see Fig \ref{plot_poweru_early}.
The convergence toward Kazantsev scaling
is seen in Fig.~\ref{mag32_comp}, where we have
plotted the magnetic energy spectrum for runs with
$\Rm$ between 120 and 540.
(The kinematic growth phase for Run~E was not available, because we
restarted Run~E from Run~D after it had already reached saturation.)

Originally, Kazantsev obtained the $k^{3/2}$ spectrum under the assumption
that the velocity has power only at large scales, which would correspond
to a large value of $\Pm$.
Simulations for large $\Pm$ have indeed confirmed the Kazantsev
slope \cite{MC01}.
Our results now show that the Kazantsev spectrum is also obtained for
$\Pm=1$.

The values of $\eta$ ($=\nu$) and $\Rm$ used for the different
runs discussed above are summarized in \Tab{Truns}.

\subsection{Approach to saturation}
\label{ApproachToSaturation}

One would naively expect that the onset of saturation happens rapidly
on a dynamical time scale.
That this is not the case in helical hydromagnetic turbulence came
originally as a surprise, but the reason is now well understood to be
a consequence of asymptotic magnetic helicity conservation.
The same argument does not apply to nonhelical turbulence.
Nevertheless, there may still be a slow-down in the saturation
behavior at small wavenumbers \cite{Schek02}.

\begin{figure}[t!]\centering\includegraphics[width=0.5\textwidth]
{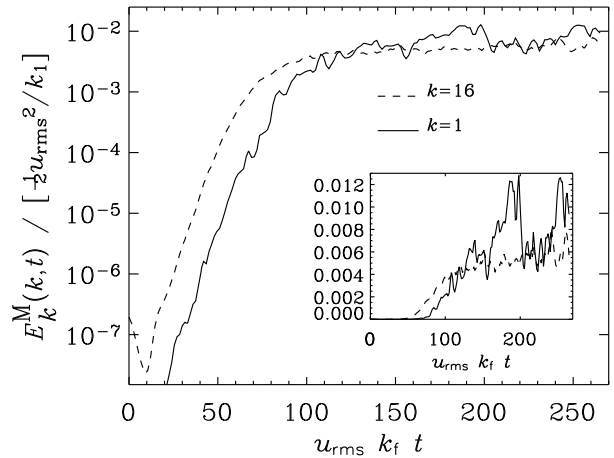}\caption{
Saturation behavior of the spectral magnetic energy at wavenumbers
$k=1$ (solid line) and $k=16$ (dashed line) for run D2.
Note the slow saturation behavior for $k=1$.
}\label{ppower_time_nohel512d2}\end{figure}

The clarification of this question is hampered by the fact that one
needs very high resolution before one can with certainty distinguish
resistive time scales from dynamical ones.
\FFig{ppower_time_nohel512d2} shows the saturation behavior for two
different values of $k$ in a run with $512^3$ meshpoints.
The initial growth is clearly exponential both at $k=1$ and at $k=16$.
However, when the magnetic energy at $k=16$ saturates at
$t\approx100/(u_{\rm rms}k_{\rm f})$, the magnetic energy at $k=1$
seems to continue growing approximately linearly \cite{Schek02};
see \Fig{ppower_time_nohel512d2} for $100<u_{\rm rms}k_{\rm f}t<200$.
Looking at the simulation at later times one sees, however, that the time
sequence is actually very bursty and that the approximately linear growth
in the interval $100<u_{\rm rms}k_{\rm f}t<200$ was only a transient.
(This is best seen in the inset of \Fig{ppower_time_nohel512d2}
where the energies are shown in a linear plot.)
We can therefore not confirm with certainty that the difference in the
saturation times for $k=1$ and $k=16$ is explained by the difference
between dynamical and resistive time scales.

In the following section we present further properties of these
runs after the time when the field has reached saturation.

\section{Dynamically saturated phase}

When the magnetic energy has reached a certain fraction of the kinetic
energy, the magnetic field stops growing exponentially and eventually reaches
a statistically steady state.
In this section we discuss the properties of this state.

\subsection{Energy spectra}

\begin{figure}[t!]\centering\includegraphics[width=0.5\textwidth]
{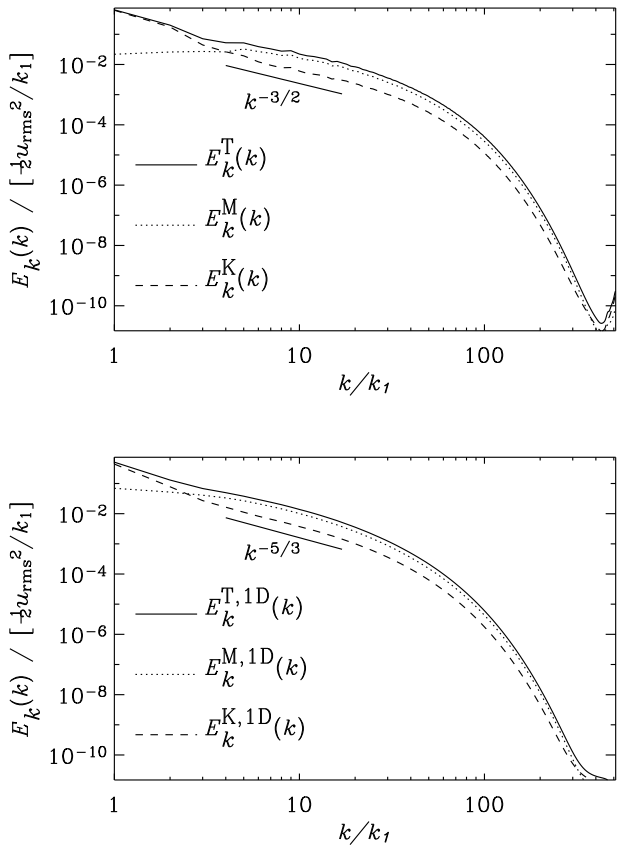}\caption{
Three-dimensional (upper panel) and 
one-dimensional (lower panel) time averaged spectra, normalized by 
$\frac{1}{2}u_{\rm rms}^2/k_1$, for Run~E with a resolution
of $1024^3$ meshpoints. The energy spectra have been averaged over a 
period of five turnover times $(u_{\rm rms}k_{\rm f})^{-1}$.
}\label{power1024a2_1d}\end{figure}

As saturation sets in, the spectral magnetic energy begins to exceed the
spectral kinetic energy at small scales;
see the upper panel of \Fig{power1024a2_1d}, where we also show the total
energy spectrum, $E^{\rm T}_k=E^{\rm K}_k+E^{\rm M}_k$.
There is a short inertial range
with an approximate Kolmogorov $k^{-5/3}$ spectrum, but, as
already reported in Paper~I and discussed in Ref.~\cite{DHYB03},
there is a strong bottleneck effect
in the three-dimensional spectra which is less strong in one-dimensional
spectra.
Therefore we plot in \Fig{power1024a2_1d} for comparison
both three-dimensional and
one-dimensional spectra for the same Run~E.
With one-dimensional
spectra we mean spectra calculated from variations along one of the 
coordinate directions only.
Throughout this paper, we refer to the one-dimensional spectra as the sum of the
longitudinal plus two times the transversal spectra \cite{DHYB03}.
To obtain the three-dimensional spectra we have integrated
over three-dimensional shells in $\kk$ space, assuming isotropy.

Much of the large-scale kinetic energy is probably
transferred directly to smaller-scale magnetic fields.
At large scales the three-dimensional magnetic energy spectrum is weakly increasing,
it peaks at $k=5$, and
then joins the $k^{-5/3}$ slope of kinetic energy, but with an approximately $2.5$ times
larger amplitude.
The one-dimensional energy spectrum, on the other hand,
is monotonically decreasing also for small wavenumbers.

All spectra terminate around the nominal dissipation cutoff wavenumber,
$k_{\rm d}=(\epsilon/\nu^3)^{1/4}$,
where $\nu=\eta$ and $\epsilon=\epsilon_{\rm K}+\epsilon_{\rm M}$ is the total
energy dissipation rate per unit mass, and
\EQ
\epsilon_{\rm K}=2\nu\bra{\rho\SSSS^2}/\rho_0\quad\mathrm{and}\quad
\epsilon_{\rm M}=\eta\mu_0\bra{\JJ^2}/\rho_0
\EN
are the contributions
from viscous and ohmic dissipation, respectively.
As the resolution is increased, the inertial range of the total energy
spectrum becomes progressively longer (see Paper~I). 
For Run~E we find $k_{\rm d}\approx142$.
The `hook' in the spectrum for $k>400$ (see \Fig{power1024a2_1d})
is probably a consequence of finite
resolution and is typical also of turbulence simulations using spectral
codes \cite{VM91}.

\begin{figure}[t!]\centering\includegraphics[width=0.5\textwidth]
{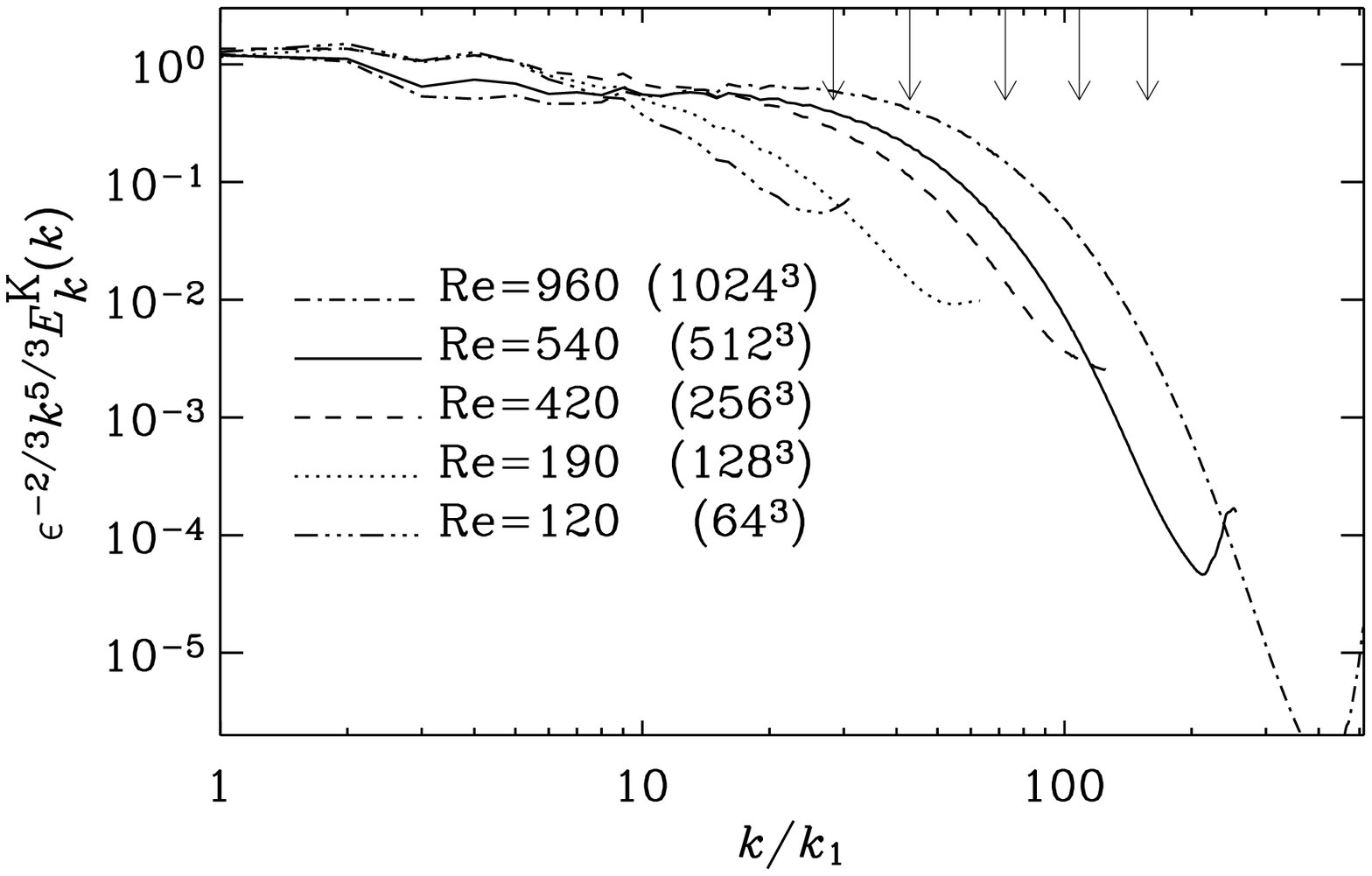}
\includegraphics[width=0.5\textwidth]
{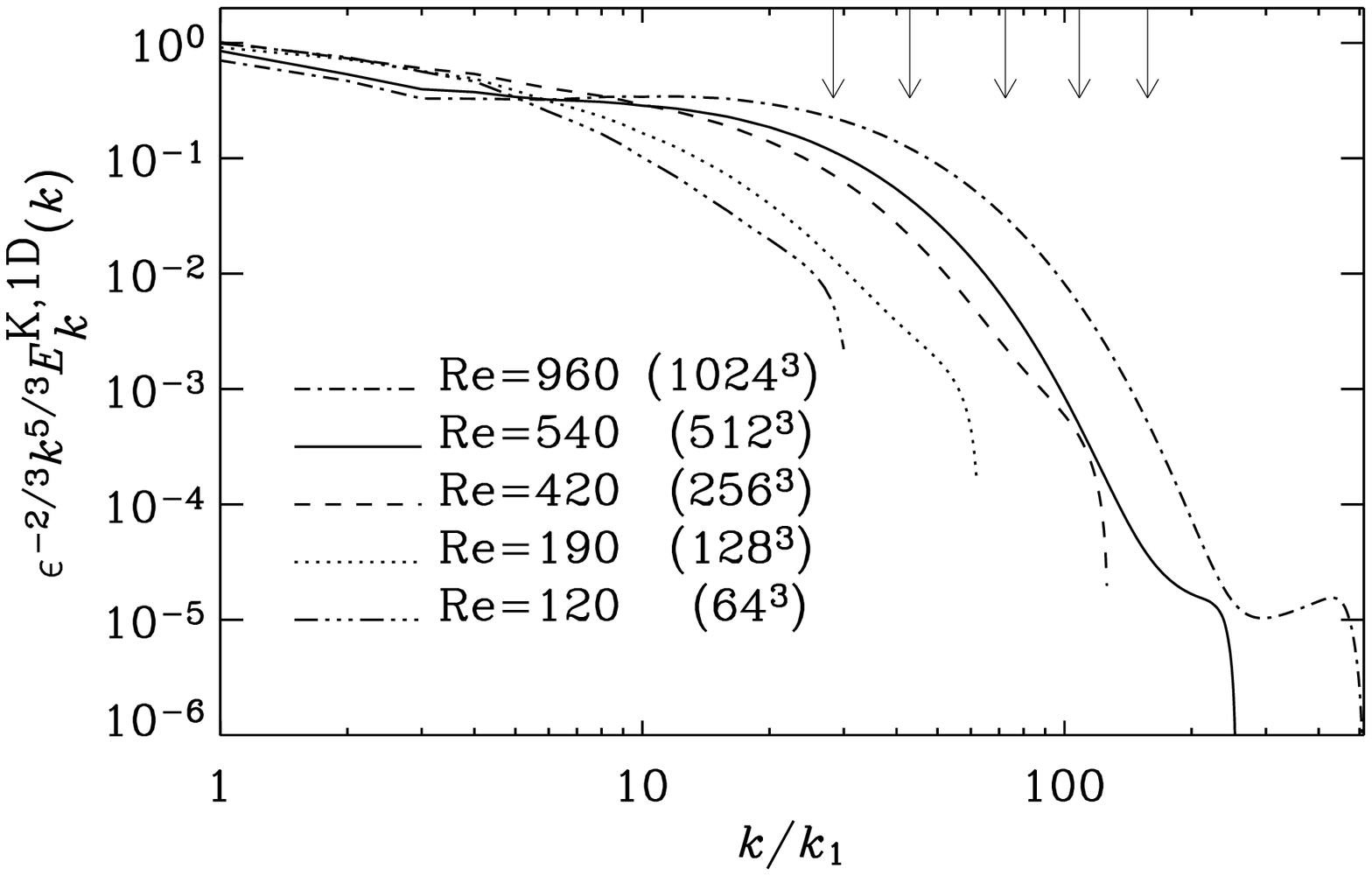}
\caption{
  Convergence of compensated three-dimensional (top) and one-dimensional
  (bottom) energy spectra of kinetic energy.
  The vertical arrows at the top show the dissipation cutoff wavenumber
  $(\epsilon/\nu^3)^{1/4}$ for the different resolutions.
}\label{kin52_comp}\end{figure}

For the numerical simulations to work and to reproduce the turbulent
cascade reliably, a certain minimum dynamic range in the compensated
spectrum $k^{5/3} E_k(k)$ is necessary.
We emphasize that this dynamic range
increases with increasing Reynolds number, as can be seen in
\Fig{kin52_comp}, where we have used viscosities close to the minimum
values required for reliable results.

\begin{figure}[t!]\centering\includegraphics[width=0.50\textwidth]
{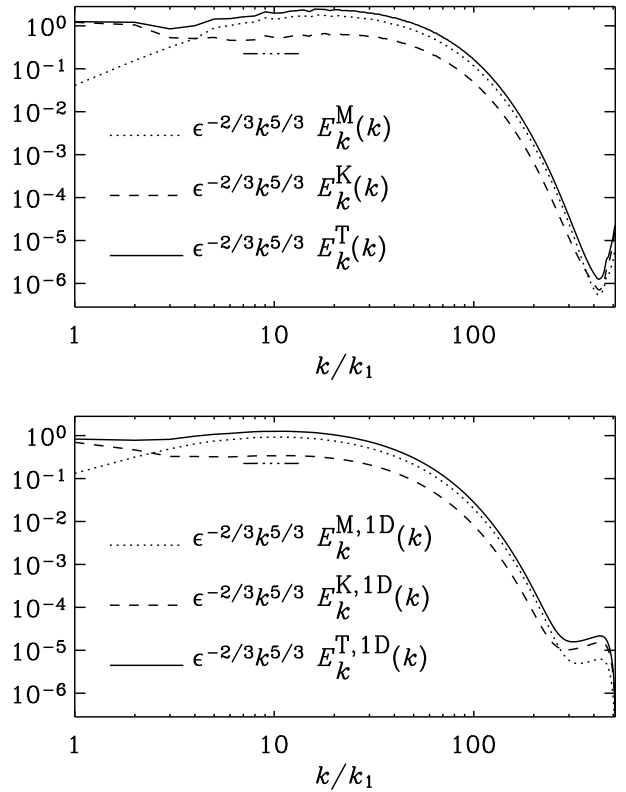}\caption{
  Compensated
  magnetic, kinetic and total three- and one-dimensional energy spectra
  for Run~E.
}\label{power1024a2}\end{figure}

Looking at one-dimensional spectra of the magnetic, kinetic and total 
energy (denoted by $E^{\rm M,1D}_k$, $E^{\rm K,1D}_k$
and $E^{\rm T,1D}_k$),
compensated with $\epsilon^{-2/3}k^{5/3}$,
one sees that around
$k=10$ there is a short range where all three compensated spectra are flat;
see \Fig{power1024a2}.
We emphasize again that this is seen only in the higher resolution runs.

There is some ambiguity as to what one calls the inertial range.
Here we refer to inertial range as only the range where the {\it total\/}
energy spectrum is compatible with a $k^{-5/3}$ slope.
The kinetic energy alone, however, can show a $k^{-5/3}$ slope
already for smaller wavenumbers; see \Fig{kin52_comp}.

\subsection{Comparison with nonmagnetic turbulence}

\begin{figure}[t!]\centering\includegraphics[width=0.5\textwidth]
{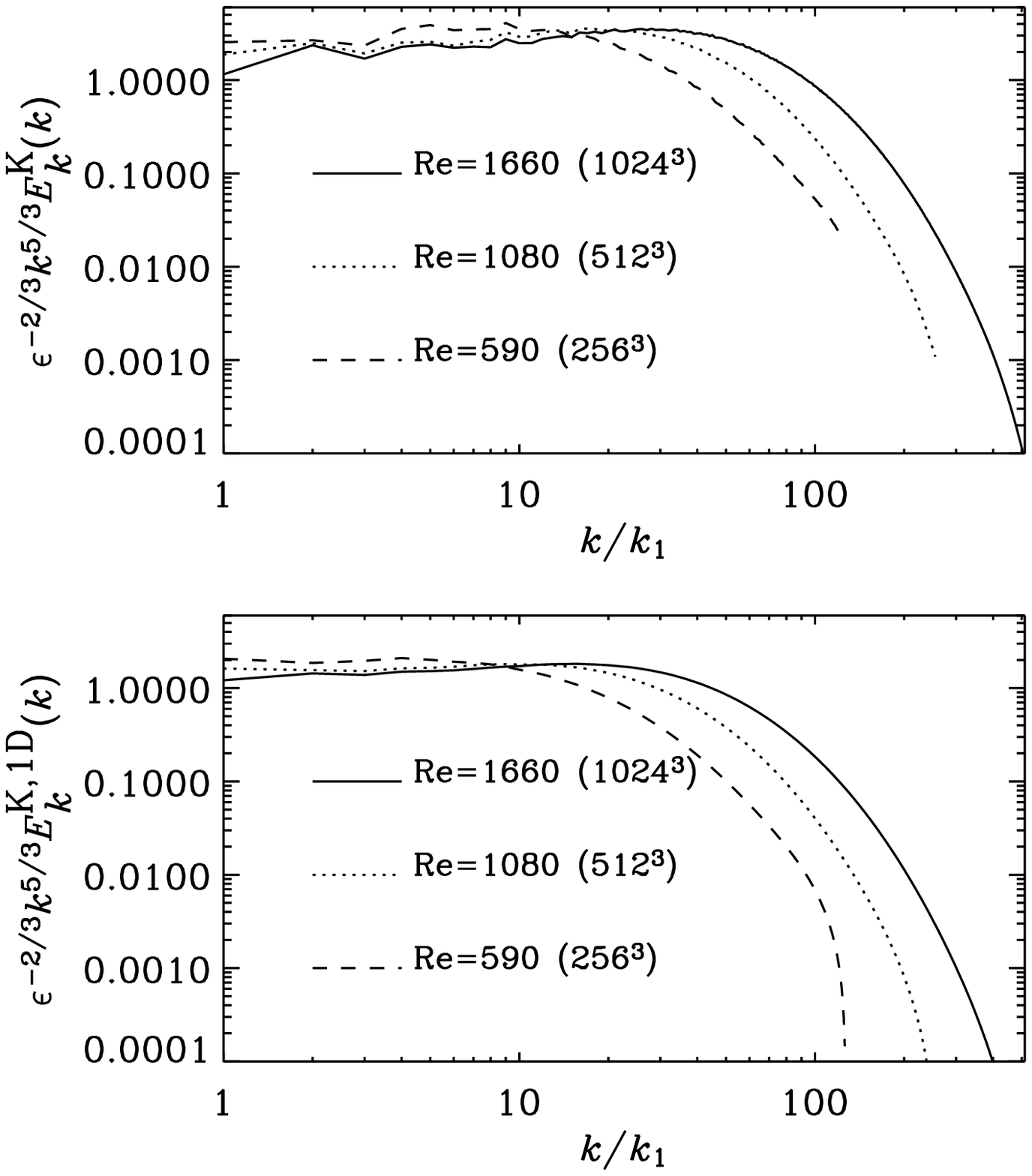}\caption{
Same as \Fig{kin52_comp}, but for a run without magnetic fields
(pure hydrodynamics), for three different values of $\Reynolds$.
Note the appearance of a bottleneck effect in the range
$10<k<50$, particularly in the 3-D spectra. For these runs $k_{\rm f}=1.5$,
and $u_{\rm rms} \approx 0.18$.
}\label{power_comp_hydro}\end{figure}

For comparison we show the resulting spectra for purely hydrodynamic
turbulence without magnetic fields and find again very similar
bottleneck behavior, as shown in \Fig{power_comp_hydro}.
A similar bottleneck effect is also found in other numerical simulations
on up to $1024^3$ meshpoints \cite{gotoh02,PWP98}.
As one increases the resolution
even further (up to $4096^3$ meshpoints)  \cite{Kan03}
the bottleneck assumes an asymptotic shape, and begins
to separate from the inertial range.
A weak bottleneck effect is found even in wind tunnel experiments
\cite{SJ93,lohse}.
In such experiments one usually measures one-dimensional longitudinal
energy spectra, and hence
a much weaker bottleneck effect is expected \cite{DHYB03}.
A physical explanation of
the bottleneck effect is found in Ref.~\cite{Fal94}.

\subsection{Subinertial range behavior}

\begin{figure}[t!]\centering\includegraphics[width=0.5\textwidth]
{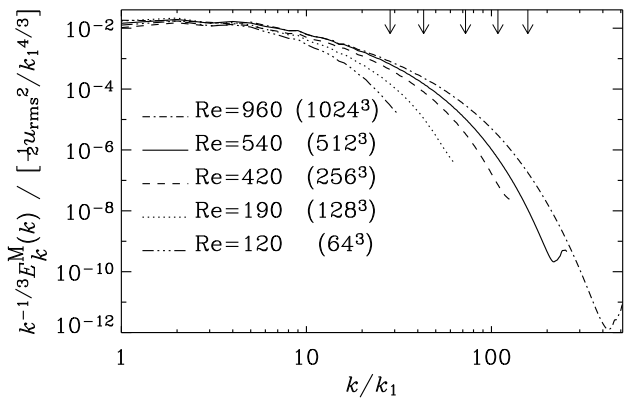}\caption{
Saturated three-dimensional magnetic energy spectra compensated by $k^{-1/3}$.
The results are comparable with the vorticity-like $k^{1/3}$ scaling
at small wavenumbers ($k<4$).
  The vertical arrows at the top show the dissipation cutoff wavenumber
  $(\epsilon/\nu^3)^{1/4}$ for the different resolutions.
}\label{mag0_comp}\end{figure}

For the saturated state, Fig.~\ref{mag0_comp} shows that
for $k \le 5$ the magnetic energy spectrum follows
approximately a $k^{1/3}$ behavior, as was originally expected
by Batchelor \cite{Bat50}.
The same slope has also been found for convective turbulence
at the time when the magnetic field is still weak \cite{BJNRST96},
and, more recently (but probably for different reasons) for ABC-flow
dynamos \cite{BDS02}.

\begin{figure}[t!]\centering\includegraphics[width=0.5\textwidth]
{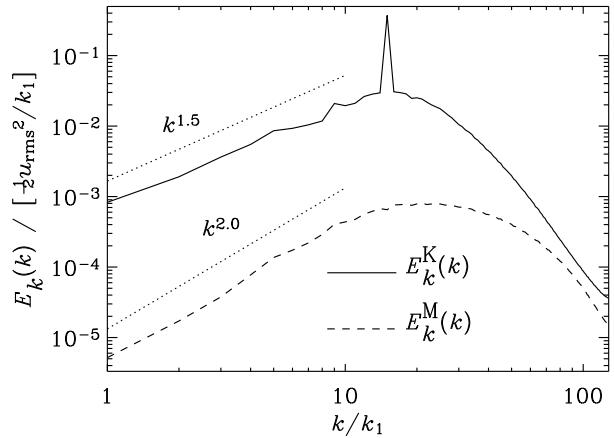}\caption{
The sub-inertial range of hydromagnetic turbulence forced at
$k_{\rm f}=15$ and normalized by $\frac{1}{2}u_{\rm rms}^2/k_1$. 
$\Rm\approx37$, $256^3$ meshpoints, $\Pm=1$.
}\label{subinertial}\end{figure}

For wavenumbers much less than the forcing wavenumber (subinertial range),
the magnetic and kinetic energy spectra increase with $k$ approximately
like $k^2$.
The theory for this spectrum is reviewed in the book by Lesieur \cite{L90}.
This scaling is best seen in spectra where the turbulence is forced at
$k_{\rm f}\gg k_1\equiv1$; see \Fig{subinertial} where $k_{\rm f}=15$.
In this figure the kinetic spectrum is somewhat shallower and scales more like
$k^{1.5}$.
In this run, the magnetic energy is saturated,
and yet it is only about 4\% of the kinetic energy.
This is mainly because the magnetic Reynolds number is only
about 37, so the run is just weakly supercritical.

\subsection{Reynolds number dependence}

\begin{figure}[t!]\centering\includegraphics[width=.5\textwidth]
{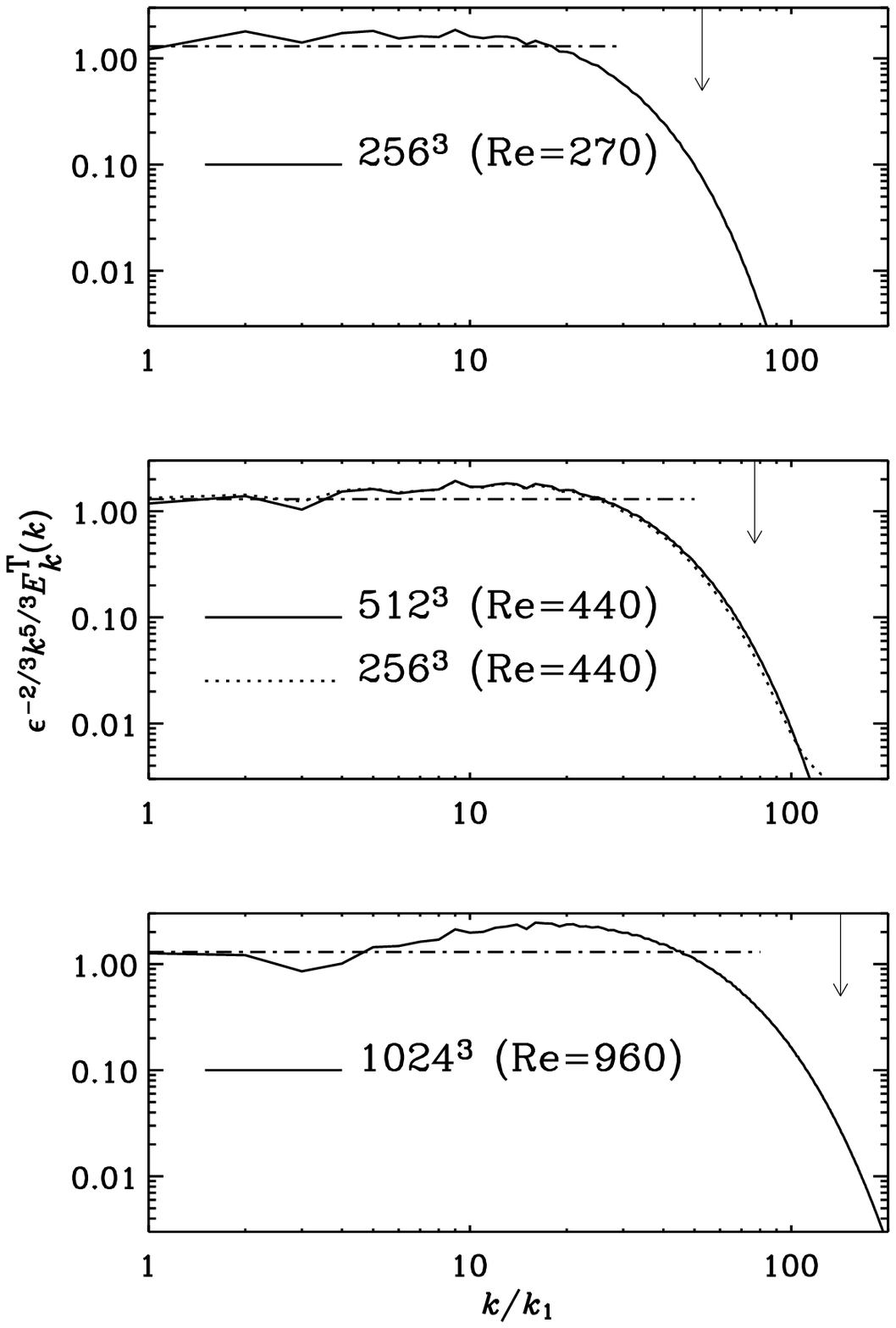}\caption{
Comparison of compensated three-dimensional total energy spectra for runs
with magnetic Prandtl number unity but with different Reynolds numbers.
In all runs the horizontal dash-dotted line represents
the value 1.3.
In the second panel two runs with the same Reynolds numbers,
but different resolution are compared.
}\label{conv_test}
\end{figure}

In \Fig{conv_test} we show total
energy spectra for three values
of $\Reynolds=\Rm$ between 270 and 960.
In the first case with $\Reynolds=\Rm=270$ the compensated
spectrum shows a reasonably flat range with 
a Kolmogorov constant $C_{\rm KYM}\approx1.3$,
which is somewhat less than the value of 2.1 found by Kida et al.\ \cite{KYM91}.
In the second case with $\Reynolds=\Rm=440$ the compensated
spectrum shows clear indications of excess power just before turning
into the dissipative subrange.
This is just the bottleneck effect and it
becomes even more dramatic in the third case with
$\Reynolds=\Rm=960$.
Thus, the bottleneck effect is only seen at sufficiently large resolution.

In order to assess the reliability of the results we have carried
out a convergence study using the same value of $\nu$, but different
mesh resolution; see the second panel of \Fig{conv_test}.
In both cases, $\nu=\eta=2\times10^{-4}$, while $u_{\rm rms}=0.13$
for $512^3$ meshpoints and 0.12 for $256^3$.
The energy dissipation is also similar, $\epsilon=2.8\times10^{-4}$
for $512^3$ meshpoints and $2.3\times10^{-4}$ for $256^3$.
The compensated energy spectra agree quite well for the two different
resolutions, and both show excess power (bottleneck effect)
just before turning into the
dissipative subrange.
This supports the conclusion that the excess power is not a numerical
artifact.

\subsection{Convergence of energy and dissipation rate}

\begin{figure}[t!]\centering\includegraphics[width=0.5\textwidth]
{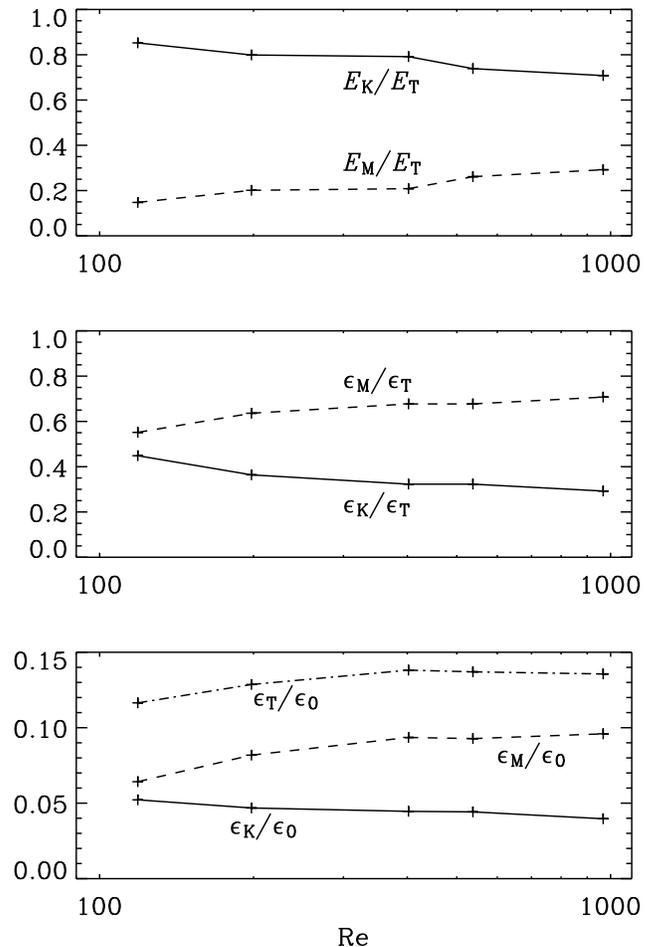}\caption{
Relative magnetic and kinetic energy (upper panel),
their respective relative dissipation rates (middle panel),
and the energy dissipation rates in units of
$\epsilon_0\equiv k_1 u_{\rm rms}^3$ (last panel),
as a function of Reynolds number for $\Pm=1$.
}\label{Fenergy_dissipation_ratio}\end{figure}

In the saturated state, the fractional magnetic and kinetic energies
tend to a constant value at large Reynolds number, with
\EQ
E_{\rm M}:E_{\rm K}\approx0.3:0.7,
\label{eratio}
\EN
so $E_{\rm M}/E_{\rm K}$ is about 0.4;
see the upper panel of \Fig{Fenergy_dissipation_ratio}. This fraction may
still depend on the forcing wavenumber since the infrared part 
of the kinetic and magnetic energy spectra can have different slopes; see
\Fig{subinertial}. Naively one would just compare $E_{\rm M}/E_{\rm K}$ for
the three non-helical simulations in \Fig{kida_test}, and find the ratio
to decrease as the forcing wavenumber is increased. This is however 
not correct since all these runs have different (magnetic) Reynolds numbers.
A more comprehensive study, where the Reynolds number is kept constant, is
therefore necessary in order to find whether or not $E_{\rm M}/E_{\rm K}$ really
does depend on the forcing wavenumber.

Equally important is the fact that also the
energy dissipation rates are converged for large
Reynolds numbers, such that
\EQ
\epsilon_{\rm M}:\epsilon_{\rm K}\approx0.7:0.3,
\label{epsratio}
\EN
i.e.\ $\epsilon_{\rm M}$ exceeds $\epsilon_{\rm K}$ by a factor of about 2.3;
see the two lower panels of \Fig{Fenergy_dissipation_ratio}.
The reciprocal correspondence of the ratios in \Eqs{eratio}{epsratio}
is coincidental.

The fact that the dissipation rates for both magnetic and kinetic energies
are asymptotically independent of Reynolds number is consistent with the basic
Kolmogorov phenomenology that leads to the scale-free $k^{-5/3}$ spectrum.
This result seems to exclude the possibility that in the large Reynolds
number limit the magnetic energy spectrum peaks at small scales. It is worth
mentioning that the ratio $\epsilon_{\rm M}/\epsilon_{\rm K}$ depends only
weakly on $\Pm$.
In \Tab{Prandtl_dep} we show that a 15 fold increase
of the magnetic Prandtl number decreases the energy dissipation ratio
only by a factor of about two; the data can be parameterized by the
power law
$\epsilon_{\rm M}/\epsilon_{\rm K}\approx2.2\;\Pm^{-1/4}$.

\begingroup
\begin{table}[t!]
\centering
\caption{
Energy dissipation rates for four runs with different magnetic Prandtl
numbers $\Pm$, showing that 
$\epsilon_{\rm M}/\epsilon_{\rm K}$ is only weakly dependent on $\Pm$.
}
\label{Prandtl_dep}
\begin{ruledtabular}
\begin{tabular}{cccccc}
$\nu\times10^4$ &$\eta\times10^4$&$\Pm$
&$\epsilon_{\rm K}\times10^4$&$\epsilon_{\rm M}\times10^4$
&$\epsilon_{\rm M}/\epsilon_{\rm K}$ \\
\hline
1.5 &  4.5  &  0.33   &   0.79 & 2.4 & 3.0 \\ 
1.5 &  1.5  &  1.     &   0.92 & 2.1 & 2.3 \\
2.0 &  2.0  &  1.     &   0.87 & 1.9 & 2.2 \\
7.5 &  1.5  &  5.     &   1.2  & 1.8 & 1.5 \\
\end{tabular}
\end{ruledtabular}
\end{table}
\endgroup

\begin{figure}[t!]\centering\includegraphics[width=0.48\textwidth]
{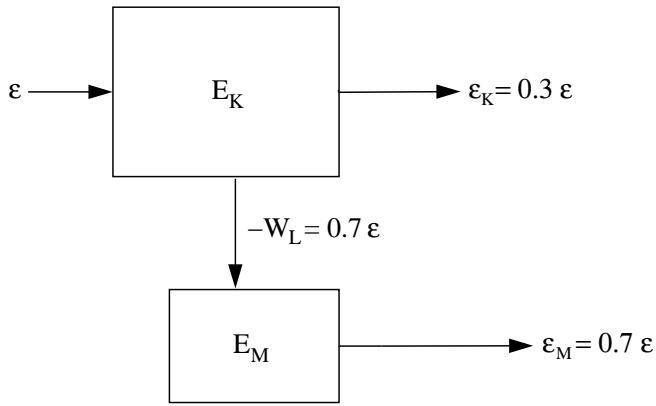}\caption{
Schematic view of the energy transport.
Most of the energy (70\%) resides in the kinetic energy reservoir, but only
30\% of the total energy input is dissipated directly by viscous heating.
Instead, 70\% of the energy flows into the magnetic energy reservoir, and
is finally dissipated by Ohmic heating.
}
\label{scheme}
\end{figure}

About 70\% of the energy that goes into the Kolmogorov cascade
is eventually converted into magnetic energy by dynamo action
and is then finally converted into heat via resistivity.
A sketch of the energy budget is given in \Fig{scheme}.

In the present simulations the thermal energy bath is not explicitly
included, but we note that in hydromagnetic turbulence simulations with shear and
rotation the resistive heat can become so important that the
temperature can increase by a factor of 10; see Ref.~\cite{BNST96}.

\subsection{Large magnetic Prandtl numbers}

\begin{figure}[t!]\centering\includegraphics[width=0.5\textwidth]
{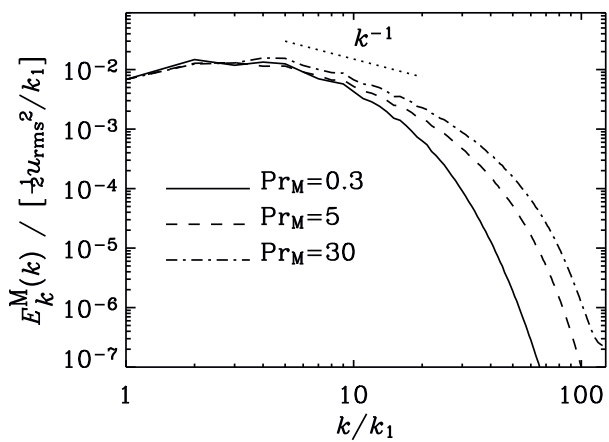}\caption{
Magnetic energy spectra for runs with magnetic Prandtl numbers ranging 
from 0.3 to 30.
}\label{prandtl256}\end{figure}

It has previously been argued that for $\Pm\gtrsim1$ the magnetic energy spectrum is
peaked at small scales \cite{MC01,MB02,SCHMMW02,Scheko02a}.
For $\Pm=1$ this claim cannot be confirmed at large Reynolds number
(see Paper~I).
The original motivation for a peak at small scales is based on linear
theory \cite{Kaz68}, which predicts a $k^{+3/2}$ spectrum; see also
\Sec{SKazantsev}.
However, the original Kazantsev model is only valid in the limit where the
velocity has only large scale components, which corresponds to $\Pm\gg1$.
In order to see how our results change with varying magnetic Prandtl
number we have calculated models for different values of $\Pm$.
One of the results is the possible emergence of a $k^{-1}$ tail in the
magnetic energy spectrum; see \Fig{prandtl256}.
The $k^{-1}$ tail has recently been found in large $\Pm$ simulations with an
imposed magnetic field \cite{CLV02b}.
The $k^{-1}$ spectrum has its roots in early
work by Batchelor \cite{Bat59} for a passive scalar and Moffatt \cite{Mof63}
for the magnetic case.

In the run with $\Pm=30$ the viscous cutoff wavenumber is
$k_\nu=(\epsilon_{\rm K}/\nu^3)^{1/4}\approx12$, so the $k^{-1}$ tail
is expected for wavenumbers larger than that.
The plot seems to suggest that the entire inertial range could
have a $k^{-1}$ spectrum, although this may well be an artifact of
an insufficient inertial range.
Instead, a more likely scenario is that for large hydrodynamic Reynolds
numbers and large magnetic Prandtl numbers there is still a $k^{-5/3}$
range for both kinetic and magnetic energies, followed by a $k^{-1}$
subrange for magnetic energy beyond the viscous cutoff wavenumber.
In any case, the peak of magnetic energy would still be at small
wavenumbers.
In summary, therefore, we find no indication of a peak of the magnetic
energy spectrum at the resistive wavenumber.

\subsection{Autocorrelation functions}

\begin{figure}[t!]\centering\includegraphics[width=0.5\textwidth]
{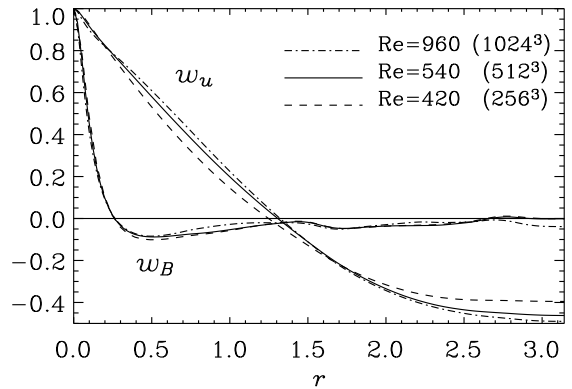}\caption{
Autocorrelation functions of magnetic field and velocity.
Note that the autocorrelation functions are nearly independent
of resolution and Reynolds number.
The velocity correlation length is $\sim3$ while
the magnetic correlation length is $\sim0.5$.
}\label{calc_correl}\end{figure}

In the context of the Zeldovich \cite{ZRS83,ZRS90} stretch-twist-fold
dynamo, the shape of the auto-correlation function of the magnetic field,
\EQ
w_B(r)=\bra{\BB(\xx)\cdot\BB(\xx{+}\rrr)}/\bra{\BB^2},
\EN
plays an important role; see also Ref.~\cite{Sub98}.
We have calculated $w_B(r)$ from the Fourier transform of the
three-dimensional, time-averaged magnetic energy spectra,
$E^{\rm M}_k(k)$; see \Fig{calc_correl}.
The diffusive scale corresponds to the thickness of the narrow
spike of $w_B(r)$ around the origin.
The typical scale over which the magnetic field changes direction
is the correlation length which corresponds to the scale where $w_B(r)$
has its minimum.
We have compared $w_B(r)$ with the velocity autocorrelation function,
$w_u(r)$, which is defined in an analogous manner.
Note that, because of isotropy, $w_u(r)$ and $w_B(r)$
are only functions of $r=|\rrr|$.
Similar autocorrelation functions have also been seen in
simulations of convective dynamos \cite{BJNRST96}.
Figure~\ref{calc_correl} shows that the velocity correlation length is
$\sim 3$ while
the magnetic field correlation length is $\sim 0.5$.
Clearly, the magnetic correlation length is much shorter than
the velocity correlation length, but it is practically independent of
$\Reynolds$ ($=\Rm$) and certainly much longer than the resistive scale,
$\sim 2\pi/k_{\rm d}\approx0.04$.

\subsection{Structure functions}

In numerical turbulence the signed structure functions for odd moments are
usually not well converged.
It is therefore customary to use unsigned structure functions,
defined as
\EQ
S_p(l)=\left< \left| \zzz(x{+}l)-\zzz(x) \right|^p \right>,
\EN
where $\zzz(x)$ is one of the two Elsasser variables 
\EQ
\zzz^\pm=\uu\pm\BB/\sqrt{\rho \mu}.
\EN
The structure function exponents $S_p(l)$ normally show a power-law
scaling
\begin{equation}
  S_p(l) \propto l^{\zeta_p},
\end{equation}
where $\zeta_p$ is the $p$th order structure function scaling exponent.
Here, because of isotropy, we usually consider the dependence on one
spatial coordinate, $x$, but we also discuss the more general case below.

In Paper~I we have shown that $\zeta_3\approx1.0$ and $\zeta_4\approx1.3$.
This is our strongest evidence that the asymptotic inertial range scaling
is $k^{-5/3}$ and that the $k^{-3/2}$ scaling seen in the upper panel of
\Fig{power1024a2_1d} is due to the bottleneck effect \cite{DHYB03}.
True Iroshnikov-Kraichnan scaling would imply $\zeta_4=1$,
whereas $\zeta_3=1$ is consistent with Goldreich-Sridhar scaling.

\begin{figure}[t!]\centering\includegraphics[width=0.5\textwidth]
{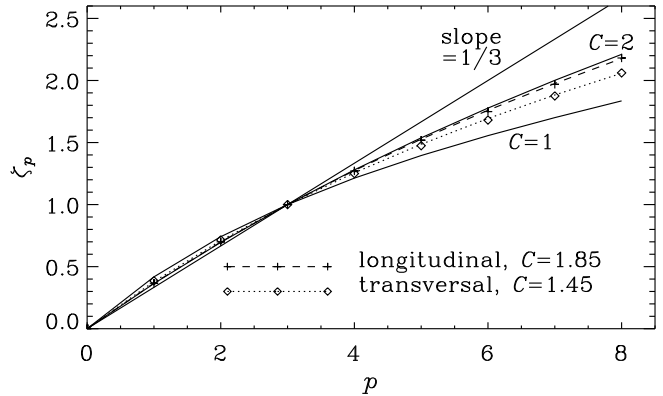}\caption{
Longitudinal and transversal structure function exponents for the 
Elsasser variables for run D.
}\label{structure-function}\end{figure}

We find that the second-order scaling exponent
is $\zeta_2=0.7$, which again indicates that the inertial range has a slope
$k^{-\zeta_2-1} \approx k^{-5/3}$, in support of our findings from 
the one-dimensional energy spectra.
Making use of the extended self similarity hypothesis \cite{Benzi93}
we find that for all $p$, the values of $\zeta_p$
are consistent with the generalized She-Leveque
formula \cite{SL94}
\EQ
\zeta_p=\frac{p}{9}+C\left[1-\left(1-\frac{2/3}{C}\right)^{p/3}\right],
\EN
where $C$ is interpreted as the codimension of the dissipative
structures.
We find that $C=2$ (corresponding to 1-dimensional, tube-like dissipative
structures) gives a reasonable fit to the longitudinal structure function
exponents; see \Fig{structure-function}.
If we allow for fractal codimensions, then $C=1.85$ gives the best fit
for the longitudinal structure function exponents and $C=1.45$ for the
transversal ones.

A similar difference between transversal and longitudinal
structure functions has previously been found \cite{gotoh02,PPW02}.
It has been argued \cite{PPW02} that this difference 
is an artifact of the forcing being in the same direction as 
the direction in which the structure functions are calculated. 
In our simulation, however, the forcing is chosen randomly in an isotropic
manner. In addition we have also performed a calculation were we average
structure functions calculated in 91 different directions distributed 
isotropically over the unit sphere. In this calculation we find again the
same difference between longitudinal and transversal structure functions.
Additional support to this result comes from the fact that 
longitudinal one-dimensional energy spectra are slightly steeper
than the transversal ones.
A possible explanation for the difference between longitudinal and transversal
structure functions has been offered by Siefert \& Peinke \cite{SP03},
who find different cascade times for longitudinal and transversal spectra.
 
As stated above, our results for the longitudinal structure functions 
of the Elsasser variables follow the generalized She-Leveque 
formula with codimension $C=2$ quite well.
It is difficult to make precise comparisons with earlier work where
often somewhat different cases are considered.
The perhaps closest comparison is possible
with the transonic hydromagnetic turbulence
simulations of Padoan et al.\ \cite{PJNB03}.
They consider only
velocity structure functions, but since they have a weak  
magnetic field, the velocity is similar to the
Elsasser variables.
They find that for small Mach numbers
the structure function scaling has codimension $C\approx2$, while for
supersonic turbulence they find $C\approx1$. For increasing Mach numbers
they see a continuous decrease from $C=2$ to $C=1$.  
This is consistent
with our result of $C=1.85$ since we have a small but finite 
Mach number, and would therefore expect a somewhat smaller value
than $C=2$, but considerably larger than $C=1$.

In another set of forced turbulence simulations,
Cho et al \cite{CLV02} find that the
velocity structure function has again codimension $C=2$. 
We emphasize that in Ref.~\cite{CLV02}
the structure functions are calculated perpendicular to the local magnetic
field, and not along the global coordinate axis.
In addition they look at 
velocity scaling while we concentrate on Elsasser variables. 
Their results can therefore not straightforwardly be compared with ours,
and the similarity is probably accidental. Indeed, it is shown in
Ref.~\cite{CLV03} that structure functions calculated perpendicular
to the local
magnetic field and those calculated along the global coordinate axis 
are {\it not} comparable.  

\begin{figure}[t!]\centering\includegraphics[width=0.5\textwidth]
{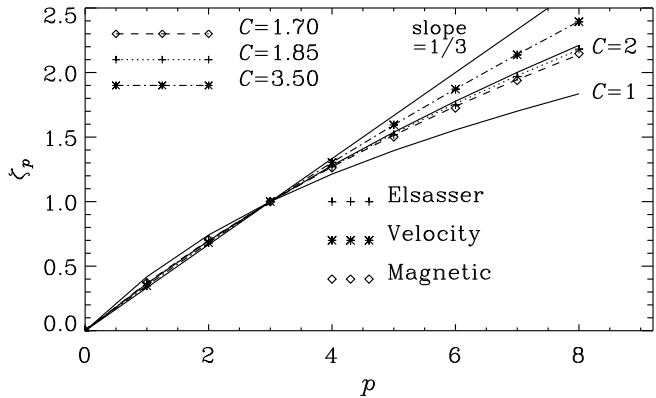}\caption{
Longitudinal structure function exponents for the Elsasser variables,
compared with those for velocity and magnetic field separately for run D.
}\label{slope_struct_ubz}\end{figure}

On the other hand, in simulations of 
decaying magnetohydrodynamic (MHD) turbulence at a resolution up to
$512^3$ collocation points,
Biskamp \& M\"uller \cite{BM00} found the
codimension of the Elsasser variables to be $C=1$. 
At first glance this seems to contradict our results, but the fact that 
they have considered decaying turbulence, 
and that the magnetic field is decaying more 
slowly than the kinetic energy, implies that they have a much stronger
magnetic field relative to kinetic field than we do. Furthermore we find
(see Fig.~\ref{slope_struct_ubz}) that the magnetic field is much more
intermittent than the velocity field \cite{CLV03}.
It is therefore not surprising
that they find different intermittency in their decay simulation than we
do in our forced simulation.

Next we look at the longitudinal structure function scaling exponents
for magnetic field and velocity, as well as for the Elsasser variables;
see Fig.~\ref{slope_struct_ubz}.
The velocity is known to be generally less intermittent than the
magnetic field \cite{CLV03}.
While the Elsasser variables follow the standard She-Leveque 
scaling ($C=2$) rather well, the velocity is less intermittent ($C=3.5$)
and the magnetic field is more intermittent ($C=1.7$).
Obviously, a codimension larger than the embedding dimension
does not have a direct geometrical meaning, so the value of $C$ for the velocity
should not be
interpreted as a codimension, but just as an indicator for a low
degree of intermittency.

\begin{figure}[t!]\centering\includegraphics[width=0.5\textwidth]
{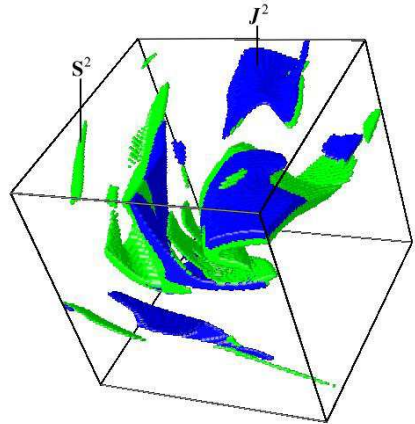}\caption{
Contours of $\JJ^2$ (blue or dark-gray)
and $\SSSS^2$ (green or light-gray) showing the
dissipative structures of magnetic and kinetic energy, respectively.
Only a very small subvolume ($1/16^3$) of the entire simulation domain
of run E is shown.
}\label{diss}\end{figure}

\begin{figure}[t!]\centering\includegraphics[width=0.5\textwidth]
{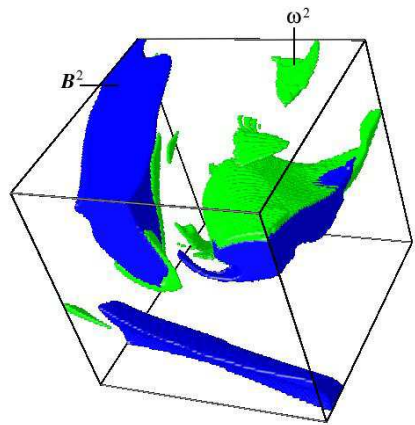}\caption{
Contours of $\BB^2$ (blue or dark-gray)
and $\oo^2$ (green or light-gray) for the same
small subvolume as shown in \Fig{diss}.
Note that the areas with strong $\BB^2$ are more extended than those with
strong $\JJ^2$ (cf.\ \Fig{diss})
and that the locations of high $\oo^2$ and high $\SSSS^2$ are near to
each other, but otherwise different in their detailed appearance
(see \Fig{diss}), even though $\bra{\oo^2}=\bra{2\SSSS^2}$.
}\label{O2_B2_structures}\end{figure}

\subsection{Visualizations}

\begin{figure*}[t!]\centering\includegraphics[width=1.0\textwidth]
{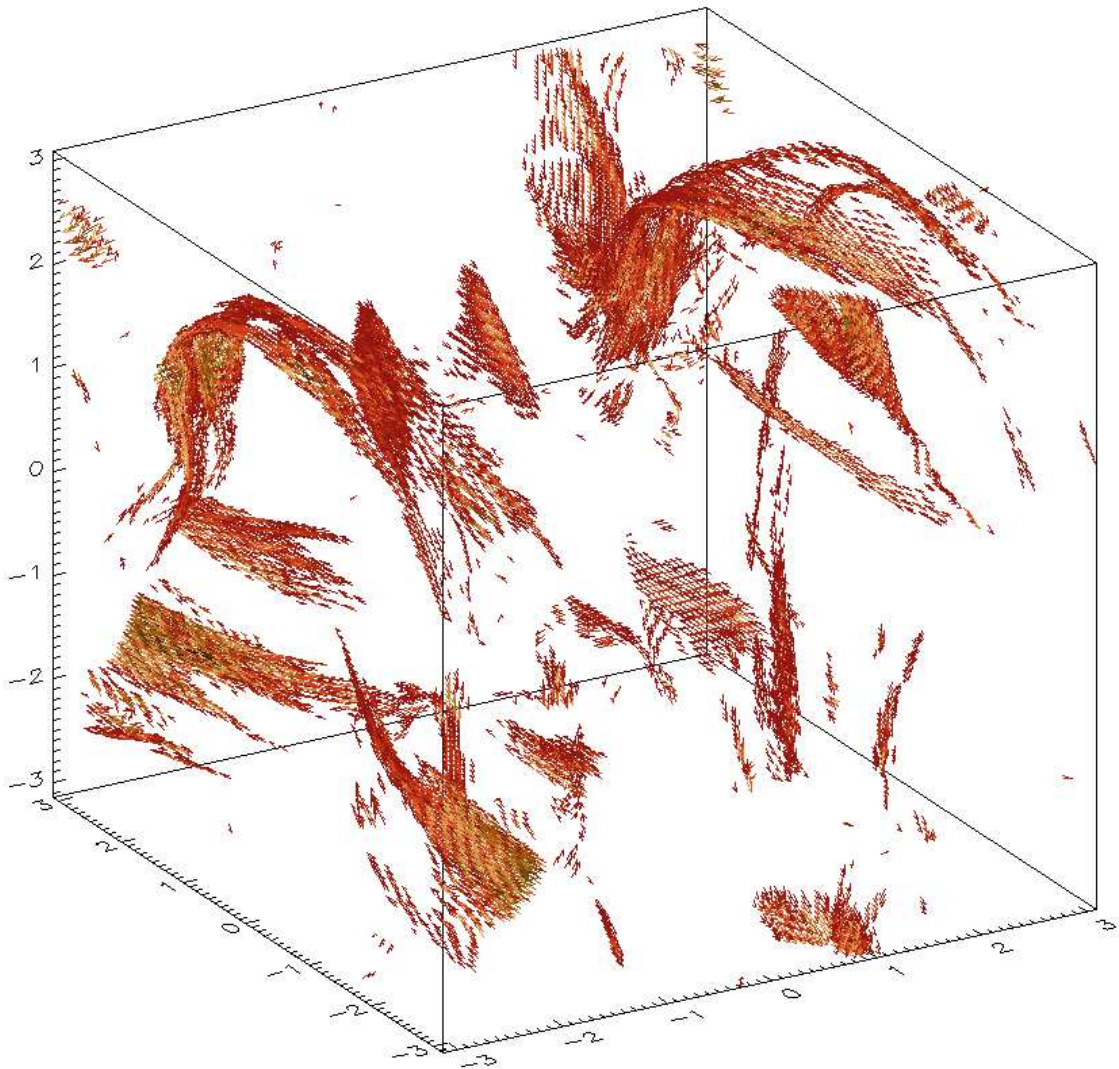}\caption{
Magnetic field vectors of run E
shown at those locations where $|\BB|>3B_{\rm rms}$.
Note the long but thin arcade-like structures extending over almost
the full domain.
The structures are sheet-like with a thickness comparable to the
resistive scale.
}\label{B_vec_3d}\end{figure*}

In view of the discussion on the dimensionality of the dissipative
structures in hydromagnetic turbulence, it is desirable to obtain an
estimate simply by visual inspection.
In \Fig{diss} we show, for a small subvolume of the
entire simulation domain, surfaces of constant $\JJ^2$ (Joule dissipation)
and $\SSSS^2$ (viscous dissipation).
Generally, the approximate dimensionality of both dissipative structures
is somewhere between sheets and tubes, although $\JJ^2$ appears to be
perhaps slightly more sheet-like.
We conclude that the dimensionality of $\JJ^2$ is consistent with
what one would have expected from the estimate $C=1.7$ obtained in
the preceding subsection.
For the dissipative structures of kinetic energy, on the other hand,
the estimate $C=3.5$ exceeded the embedding dimension and did therefore
not make geometrical sense anyway.
The qualitative inspection of $\SSSS^2$ would have suggested a codimension
between 1 and 2.
We therefore conclude that $C$ can only be regarded as a fit parameters
and that there is not always an obvious connection with the actual
appearance of the dissipative structures.

It should be noted that when one normally talks about structures in
turbulence one often talks about vortex tubes and, in the magnetic case,
magnetic flux tubes.
These are quite distinct from the dissipative structures.
Vorticity $\oo=\nab\times\uu$ and rate of strain tensor $\SSSS$
characterize respectively the
antisymmetric and symmetric parts of the velocity gradient matrix,
and are therefore not expected to look similar.
On the other hand, $\bra{\oo^2}=\bra{2\SSSS^2}$, where 
and angular brackets denote volume averages, and \Fig{O2_B2_structures}
shows that both $\oo^2$ and $\SSSS^2$ exhibit similar length scales.
The difference is more pronounced in the magnetic case, where $\JJ$
is clearly
dominated by smaller scale structures while $\BB$ can exhibit structures
of much larger scale.
This is shown in Fig.~\ref{B_vec_3d} where we visualize magnetic field vectors in the
full box at those locations where the field exceeds three times the
rms value.
The strong field turns out to be of surprisingly large scale,
even though this dynamo has no helicity and no large scale field in
the usual sense; cf.\ Ref.~\cite{B01}.
The large scale structures we find are reminiscent of the
ropes discussed in Refs~\cite{ZRS90,Sub98}, although
they seem to be more sheet-like \cite{PPS95}, and their
thickness is of the order of the resistive scale \cite{BPS95}.

\section{Connection with helical turbulence}

\begin{figure}[t!]\centering\includegraphics[width=0.5\textwidth]
{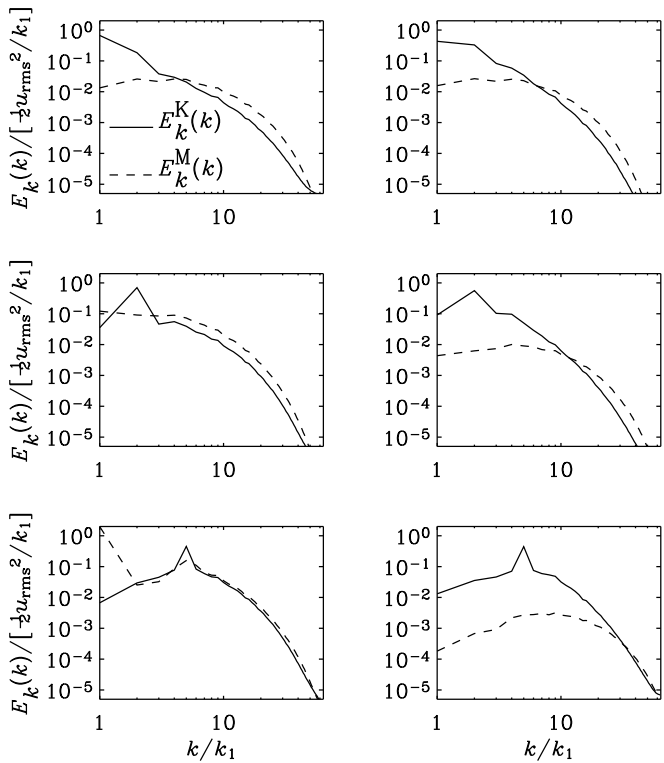}\caption{
  Three-dimensional kinetic and magnetic energy spectra, normalized by
  $\frac{1}{2}u_{\rm rms}^2/k_1$, for
  runs with and without helicity
  (left and right hand columns, respectively) and for
  the different forcing wavenumbers:
  $k_{\rm f}=1.5$ (top) $2.3$ (middle), $5.1$ (bottom).
  $128^3$ meshpoints.
}\label{kida_test}
\end{figure}

Finally we comment on some
differences between helical and nonhelical turbulence.
In the case of helical forcing one expects an {\it inverse} cascade to
smaller wavenumbers, rather than a direct cascade to larger wavenumbers.
We can now identify two reasons why this has not really been seen in early
turbulence simulations with helical forcing \cite{KYM91}.
On the one hand the inverse cascade takes a resistive
time to develop \cite{B01},
and this time tends to be too long
if magnetic hyperdiffusivity is used \cite{BS02}.
In Ref.~\cite{KYM91} magnetic hyperdiffusivity was indeed used and the resistive
time was at least a hundred times longer than the duration of the
runs, so no inverse cascade should be expected.
But there is another more important reason.
In order for the inverse cascade to develop, one has to have some scale
separation, i.e.\ the magnetic field must be allowed to grow on
scales larger than the forcing scale
(which corresponds to the energy carrying scale) of
the turbulence.
This was not the case in the early simulations and
may explain why the inverse cascade has not been seen in Ref.~\cite{KYM91} and
that those results should therefore be closer to the case without helicity.
To substantiate this, we have carried out simulations with helical
and nonhelical forcing using the modified forcing function
\begin{equation}
\ff_{\kk}=\RRRR\cdot\ff_{\kk}^{\rm(nohel)}\quad\text{with}\quad
{\sf R}_{ij}={\delta_{ij}-\ii\sigma\epsilon_{ijk}\hat{k}_k
\over\sqrt{1+\sigma^2}},
\end{equation}
where $\ff_{\kk}^{\rm(nohel)}$ is the non-helical forcing function
of \Eq{nohel_forcing}.
In the helical case ($\sigma=\pm1$) we recover the forcing function used in
Ref.~\cite{B01}, and in the non-helical case ($\sigma=0$) this forcing
function becomes equivalent to that of \Eq{nohel_forcing}.

We show in \Fig{kida_test} the
energy spectra of helical and nonhelical simulations with forcing at
wavenumber $k_{\rm f}=1.5$ (no scale separation), $k_{\rm f}=2.3$
(weak scale separation), and $k_{\rm f}=5$ (considerable scale separation).
For $k_{\rm f}=1.5$ the spectra of the helical and nonhelical simulations
are indeed quite similar to each other (e.g., no inverse cascade and slight
super-equipartition at $k>5$).
For $k_{\rm f}=5.1$, on the other hand, the spectra are quite different
and there is no inverse cascade in the nonhelical case.

\section{Conclusions}

In the present paper we have studied non-helical MHD
turbulence without imposed large scale fields. We find that, in order
to get dynamo action, the magnetic Reynolds number has to exceed the critical
value $\Rm^{\rm(crit)} \approx 30$. When the dynamo is in the kinematic 
regime (the magnetic field is still too weak to affect the velocity) we 
see a kinetic energy spectrum with the Kolmogorov $k^{-5/3}$ inertial range,
while the magnetic energy spectrum shows the expected $k^{3/2}$ Kazantsev
slope ({\it increasing} with $k$).

As the dynamo gets saturated we find that there is a short
inertial range where the magnetic and kinetic energy spectra are parallel.
This is only seen in the largest of our simulations with $1024^3$ 
meshpoints.
The magnetic energy spectrum exceeds the kinetic 
energy spectrum by a factor of $\approx 2.5$,
which seems to be more or less the asymptotic value as $\Reynolds$ grows larger.
At first glance one is led to believe that the saturated energy spectra exhibit 
a $k^{-3/2}$ inertial range, but we argue that this is due to a 
strong bottleneck effect.
For one-dimensional spectra the bottleneck effect becomes
much weaker and they have the expected $k^{-5/3}$ slope.
We have demonstrated
that simulations without magnetic fields show the same bottleneck effect.
Also the second order structure functions are consistent with
a $k^{-5/3}$ scaling.
We therefore conjecture that for larger values of $\Reynolds$ one will see
a $k^{-5/3}$ subrange also for the three-dimensional MHD spectra,
although the bottleneck effect will continue to affect at least one
decade or more in wavenumbers just before the dissipative subrange

At large scales ($1 \le k \le 5$) we see a $k^{1/3}$ behavior of the 
magnetic field.
However, if we force the flow at $k_{\rm f} \gg k_1$ (i.e.\ the energy is
being injected at scales much smaller than the box size), 
we find that for $k \ll k_{\rm f}$ the kinetic and magnetic energy
spectra scale almost like $k^2$, although the kinetic 
energy spectrum is somewhat shallower which may be an artifact
of the finite size of the computational domain.

Concerning the structure functions we find that our simulations are in 
good agreement with those found by Padoan et al.\ \cite{PJNB03}.
The Elsasser variables
follow the She-Leveque scaling with a codimension somewhat less than 2, 
which is what one would
expect since our simulation is weakly compressible. We also find that the 
magnetic field is more intermittent than the velocity, which is
qualitatively consistent with earlier findings \cite{MFP81}.
Quantitatively, in terms of structure function exponents,
this has recently also been
found by Cho et al.\ \cite{CLV03}, but it is still not known what
is the cause of this difference between the intermittency of magnetic and
kinetic fields.

In the case of magnetic Prandtl numbers larger than unity, there are
indications of a $k^{-1}$ range for the magnetic energy spectrum below the 
viscous cutoff wavenumber. In order to ensure that this is really the asymptotic 
slope, yet larger simulations are required.

\acknowledgments
We thank Ben Chandran for pointing out an error in one of the
plots of an earlier version of the paper.
We acknowledge Scientific Computing for granting time on the Horseshoe cluster,
and the Norwegian High Performance Computing Consortium (NOTUR) 
for granting time on the parallel computers in 
Trondheim (Gridur/Embla) and Bergen (Fire).


\end{document}